\documentclass{emulateapj}
\usepackage{color,epsfig,graphics,graphicx,pdfpages}
\definecolor{brown}{rgb}{0.6,0.4,0.2}
\definecolor{purple}{rgb}{0.5,0,0.5}

\def\msol{\hbox{\kern 0.20em $M_\odot$}}
\newcommand{\lsol}{\hbox{\kern 0.20em $L_\odot$}}
\newcommand{\g}{\hbox{\kern 0.20em g}}
\newcommand{\gmu}{\hbox{\kern 0.20em g$^{-1}$}}
\newcommand{\kg}{\hbox{\kern 0.20em kg}}
\newcommand{\pc}{\hbox{\kern 0.20em pc}}
\newcommand{\mum}{\hbox{\kern 0.20em $\mu$m}}
\newcommand{\mumd}{\hbox{\kern 0.20em $\mu$m$^{-2}$}}
\newcommand{\cm}{\hbox{\kern 0.20em cm}}
\newcommand{\m}{\hbox{\kern 0.20em m}}
\newcommand{\km}{\hbox{\kern 0.20em km}}
\newcommand{\nm}{\hbox{\kern 0.20em nm}}
\newcommand{\s}{\hbox{\kern 0.20em s}}
\newcommand{\h}{\hbox{\kern 0.20em h}}
\newcommand{\smu}{\hbox{\kern 0.20em s$^{-1}$}}
\newcommand{\smd}{\hbox{\kern 0.20em s$^{-2}$}}
\newcommand{\an}{\hbox{\kern 0.20em an}}
\newcommand{\anmu}{\hbox{\kern 0.20em an$^{-1}$}}
\newcommand{\yr}{\hbox{\kern 0.20em yr}}
\newcommand{\yrmu}{\hbox{\kern 0.20em yr$^{-1}$}}
\newcommand{\Myr}{\hbox{\kern 0.20em Myr}}
\newcommand{\Mymu}{\hbox{\kern 0.20em Myr$^{-1}$}}
\newcommand{\K}{\hbox{\kern 0.20em K}}
\newcommand{\pcmu}{\hbox{\kern 0.20em pc$^{-1}$}}
\newcommand{\pcmd}{\hbox{\kern 0.20em pc$^{-2}$}}
\newcommand{\pcmt}{\hbox{\kern 0.20em pc$^{-3}$}}
\newcommand{\kms}{\hbox{\kern 0.20em km\kern 0.20em s$^{-1}$}}
\newcommand{\kmpd}{\hbox{\kern 0.20em km$^{2}$}}
\newcommand{\kpc}{\hbox{\kern 0.20em kpc}}
\newcommand{\cms}{\hbox{\kern 0.20em cm\kern 0.20em s$^{-1}$}}
\newcommand{\erg}{\hbox{\kern 0.20em erg}}
\newcommand{\ergs}{\hbox{\kern 0.20em erg}}
\newcommand{\cmpd}{\hbox{\kern 0.20em cm$^2$}}
\newcommand{\cmmd}{\hbox{\kern 0.20em cm$^{-2}$}}
\newcommand{\cmms}{\hbox{\kern 0.20em cm$^{-6}$}}
\newcommand{\cmpt}{\hbox{\kern 0.20em cm$^3$}}
\newcommand{\cmmt}{\hbox{\kern 0.20em cm$^{-3}$}}
\newcommand{\mpd}{\hbox{\kern 0.20em m$^2$}}
\newcommand{\mmd}{\hbox{\kern 0.20em m$^{-2}$}}
\newcommand{\mpt}{\hbox{\kern 0.20em m$^3$}}
\newcommand{\mmt}{\hbox{\kern 0.20em m$^{-3}$}}
\newcommand{\mujy}{\hbox{\kern 0.20em $\mu$Jy}}
\newcommand{\mjy}{\hbox{\kern 0.20em mJy}}
\newcommand{\Mj}{\hbox{\kern 0.20em MJy}}
\newcommand{\jy}{\hbox{\kern 0.20em Jy}}
\newcommand{\ghz}{\hbox{\kern 0.20em GHz}}
\newcommand{\G}{\hbox{\kern 0.20em G}}
\newcommand{\muG}{\hbox{\kern 0.20em $\mu$G}}

\newcommand{\etal}{{et al.\/~}}

\newcommand{\spitzer}{\textit{Spitzer}}

\shorttitle{Shocked Gas in G357.7+0.3}
\shortauthors{Rho et al.}

\begin{document}

\title{Discovery of Broad Molecular lines and 
of Shocked Molecular Hydrogen from the Supernova Remnant
G357.7+0.3: HHSMT, APEX, {\it Spitzer} and SOFIA Observations}

\author{
J. Rho\altaffilmark{1,2},
J. W. Hewitt\altaffilmark{3,4},
J. Bieging\altaffilmark{5},
W. T. Reach\altaffilmark{6},
M. Andersen\altaffilmark{7},
and R. G\"{u}sten\altaffilmark{8}
}
\altaffiltext{1}{SETI Institute, 189 N. Bernardo Ave., Mountain View, CA 94043; jrho@seti.org} 
\altaffiltext{2}{SOFIA Science Center,
NASA Ames Research Center, MS211-1, Moffett Field, CA 94043}
\altaffiltext{3}{CRESST/University of Maryland, Baltimore County, Baltimore, MD 21250 and
NASA Goddard Space Flight Center, Greenbelt, MD 20771, USA}
\altaffiltext{4}{University of North Florida, Dept. of Physics, Jacksonville, FL 32224; john.w.hewitt@unf.edu}
\altaffiltext{5}{Steward Observatory, The University of Arizona, Tucson AZ 85721, USA; jbieging@as.arizona.edu}
\altaffiltext{6}{Universities Space Research Association, SOFIA Science Center,
NASA Ames Research Center, MS 232, Moffett Field, CA 94034; wreach@sofia.usra.edu}
\altaffiltext{7}{Gemini Observatory, Casilla 603, La Serena, Chile, manderse@gemini.edu}
\altaffiltext{8}{Max Planck Institut f\"{u}r Radioastronomie, Auf dem Hügel 69, 53121 Bonn, Germany, guesten@mpifr-bonn.mpg.de}

\centerline{\it ApJ, in press}

\begin{abstract}

We report a discovery of shocked gas from the supernova remnant (SNR)
G357.7+0.3. Our millimeter and submillimeter observations reveal broad
molecular lines of CO(2-1), CO(3-2), CO(4-3), $^{13}$CO (2-1) and $^{13}$CO
(3-2), HCO$^+$ and HCN using HHSMT, Arizona 12-Meter Telescope, APEX and
MOPRA Telescope. The widths of the broad lines are 15-30 \kms, and the
detection of such broad lines is unambiguous, dynamic evidence showing that
the SNR G357.7+0.3 is interacting with molecular clouds. The broad lines
appear in extended regions ($>$4.5$'$$\times$5$'$). We also present
detection of shocked H$_2$ emission in mid-infrared but lacking ionic lines
using the {\it Spitzer} IRS observations to map a few arcmin area. The
H$_2$ excitation diagram shows a best-fit with a two-temperature LTE model
with the temperatures of $\sim$200 and 660 K. We observed [C~II] at
158$\mu$m and high-J CO(11-10) with the GREAT on SOFIA. The GREAT spectrum
of [C~II], a 3$\sigma$ detection, shows a broad line profile with a width
of 15.7 km $^{-1}$ that is similar to those of broad CO molecular lines.
The line width of [C~II] implies that ionic lines can come from a
low-velocity C-shock. Comparison of H$_2$ emission with shock models shows
that a combination of two C-shock models is favored over a combination of
C- and J-shocks or a single shock. We estimate the CO density, column
density, and temperature using a RADEX model. The best-fit model with
n(H$_2$) = 1.7$\times$10$^{4}$ cm$^{-3}$, N(CO) = 5.6$\times$10$^{16}$
cm$^{-2}$, and T = 75 K  can reproduce the observed millimeter CO
brightnesses.

\end{abstract}


\section{Introduction}

Supernovae (SNe) are among the most violent events in the Universe, ejecting gas 
and returning material from dense molecular clouds into the more diffuse
interstellar medium and the galactic halo.  When the expanding blast wave encounters a
dense molecular cloud, the SN shocks drive  excitation, chemical reaction, and dynamic
motion of the gas, and destroy dust grains by collisions or thermal sputtering (Jones
\etal\ 1994; Andersen \etal\ 2011). This interaction may determine the fate of the
molecular cloud, either dispersing it or triggering collapse of dense cores leading to 
a subsequent generation of star formation.

A few signs of molecular clouds and SN (MC-SN or MC-SNR) interactions have
been found such as {\it Mixed-morphology} SNRs, or detections of OH masers
and molecular hydrogen (H$_2$). A signspot of the interaction is
center-filled, thermal X-ray emission \citep[e.g.][and references
therein]{rho98, pannuti14} that can be produced through thermal conduction
\citep{tilley06a,tilley06b, orlando09}. However, it is still not clear if
the interaction with clouds is a unique mechanism to produce
Mixed-morphology SNRs. A better indicator of SN-MC interactions is the
presence of {\it 1720 MHz OH masers} which are detected from $\sim$20 SNRs
\citep{frail96, yusef-zadeh99, hewitt08}. These masers are thought to be
collisionally excited and they suggest SN-MC interactions, but maser
emission is not well understood \citep{yusef-zadeh99}. In 3C391, a maser
position in the southwestern shell is correlated with broad CO lines, but a
maser position in the northeastern shell show only narrow CO lines
\citep{reach99}. Other evidence of shock interaction with clouds is {\it
H$_2$ emission} from collisionally excited shocked gas, but the H$_2$
emission could also originate from UV pumping \citep{burton92}. Examples of
collisionally excited H$_2$ emission from shock interaction with molecular
clouds are IC 443 \citep{burton90, richter95}, W44 and W28
\citep{reach05,neufeld07}. A one-to-one correspondence between H$_2$ and
broad CO emission has been  found in IC 443 \citep{rho01}. Eighteen
interacting remnants are found using infrared (IR) colors from the
\spitzer\ GLIMPSE data (Reach \etal\ 2006). Follow-up \spitzer\
spectroscopy confirms detection of H$_2$ lines as well as ionic
fine-structure lines and shock-processed dust (Hewitt \etal\ 2009; Andersen
\etal\ 2010).

A strong correlation between molecular interacting SNRs and $\gamma$-ray
emission has been known since the EGRET era \citep{esposito96}.
Recently Fermi and HESS observations revealed extended $\gamma$-ray
emission associated with molecular interacting SNRs like IC 443
\citep{abdo10a} , W44 \citep{uchiyama10, abdo10b} and W28
\citep{hanabata14, aharonian08, abdo10c},
emphasizing the astrochemical processes of molecules and hadronic particles
by SNe. The $\gamma$-ray emission is associated with shocked molecular
material previously identified with millimeter (mm) CO and infrared H$_2$
emission (Burton et al. 1990; Reach \& Rho 2006; Reach, Rho, \& Jarrett
2005). With {\it Fermi} observations, there is a growing number of
$\gamma$-ray emitting SNRs, many of which have the indicators of MC-SNR
interactions described above \citep[][and references therein]{abdo09,
hewitt12, wu11, daniel10, hewitt15}. The study of MC-SNR interactions is
advancing rapidly because of multi-wavelength observations; strong
correlations among them provide opportunities to discover true samples of
SNRs interacting with clouds.

The clearest evidence for interaction between SNRs and molecular clouds is
the detection of emission from the shocked molecules themselves. 
Millimeter observations provide direct, unambiguous evidence of interaction
when a broad ($>$10 \kms) line caused by dynamic motion of shocked gas is
detected. There has been a long-term effort to search for interactions with
clouds using millimeter observations \citep[]{huang86, jeong13,
zhou14,liszt09}. However, detection of broad CO lines is still limited to a
half dozen SNRs. Millimeter observations of SNRs which have evidence of
shocks by showing broad lines are IC 443 \citep[e.g.][references
therein]{vandishoeck93}, W44 \citep{wootten77, reach05, seta04, anderl14},
W28 \citep{arikawa99, reach05}, 3C391 \citep{reach99}, W51C \citep{koo97},
and HB21 \citep{koo01, shinn10}. For some cases, changes in CO velocity
profiles or a small amount of broadening in the line profiles may indicate
interactions with clouds \citep{dubner04, zhou11}. Recently
\citet{kilpatrick16, kilpatrick14} observed SNRs emitting $\gamma$-ray
emission using HESS and detected evidence of interaction in millimeter from
a few SNRs.
Another example
of broad molecular lines is a water line at 557 GHz 
from the SNR G349.7+0.2
(with a FWHM of 144 km s$^{-1}$) 
using Herschel HIFI observations \citep{rho15}. IC
443 is another case showing a broad infrared water line \citep{snell05}.

\begin{figure*}
\includegraphics[scale=0.6,width=16.5truecm]{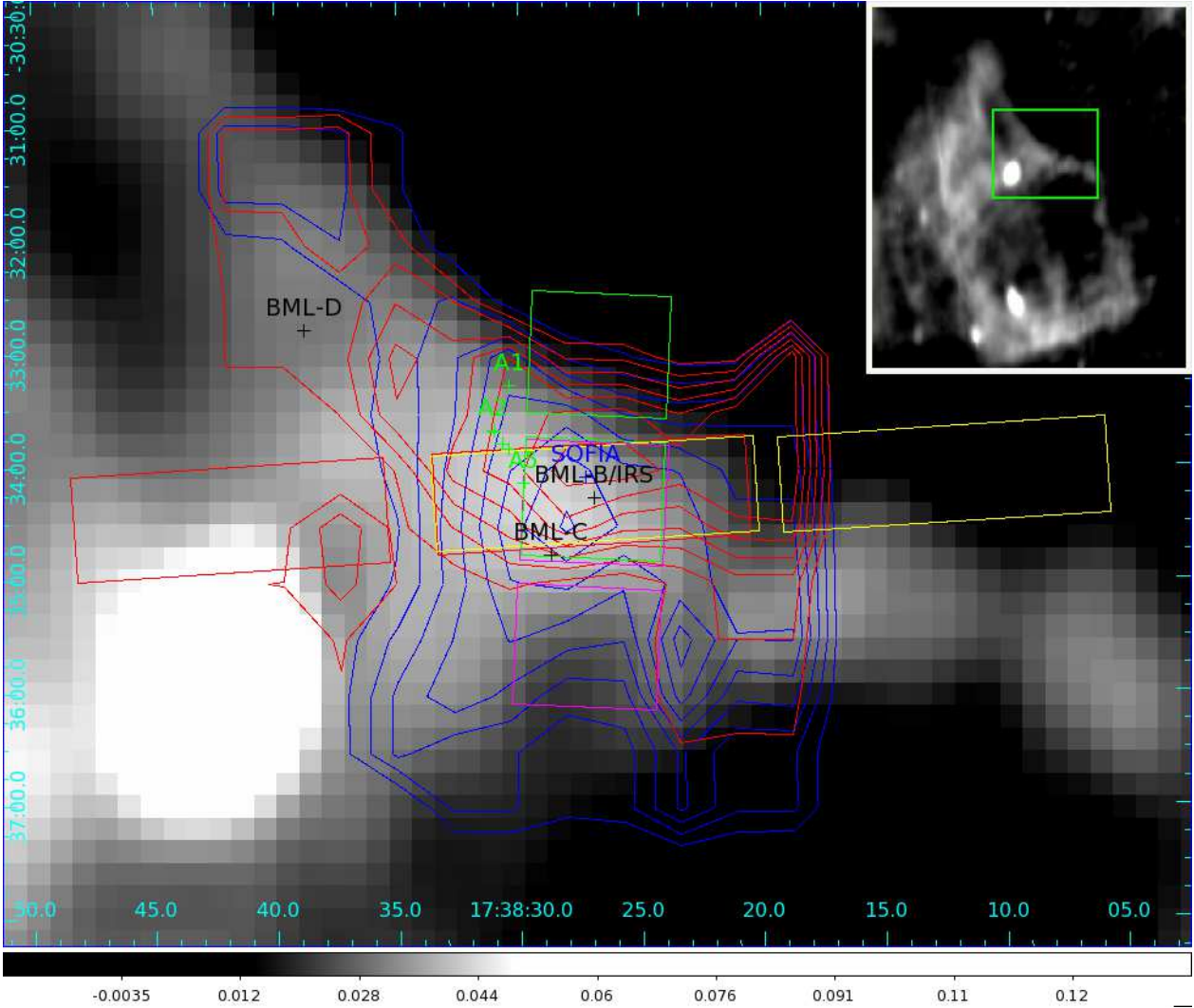}
\caption {The positions of OH masers from \citet[][crosses in green]{yusef-zadeh99}, 
CO broad molecular lines (BML; crosses in black), and the center (the same
as BML-B) and fields of view (marked in rectangular boxes) of  {\it
Spitzer} IRS (in green for SL1, magenta for SL2, red for LL2, and yellow
for LL1) are marked on the northwestern part of radio image of G357.7+0.3.
The field of view is shown in the radio image of the entire SNR (in the
inserted image marked as a box in green). The contours are from CO braod
molecular line maps (blue-wing in blue and red-wing in red; see Section 3.2
for details). The 20 cm radio continuum image shows 32$'$$\times$35$'$
field of view (FOV) centered on R.A. $17^{\rm h} 38^{\rm m} 36.5^{\rm s}$ and
Decl.\ $-30^\circ$37$^{\prime} 26.2^{\prime \prime}$ (J2000), and the scale
of color bar has a unit of Jy/beam (synthesized beam of 15$''$). }
\label{g357radio}
\end{figure*}

The SNR G357.7+0.3 is relatively unknown and under-studied. Radio observations
identified this source as a SNR \citep{reich84}. It is named $``$Square Nebula"
because of its square-like radio morphology as shown in Figure \ref{g357radio}. Soft,
faint blobs of X-ray emission using {\it Einstein IPC} have been detected with an
inferred temperature of 5.4$\times$10$^{6}$ K and an age of 10,000 yr
\citep{leahy89}. OH masers have been detected around -35 \kms\ \citep{yusef-zadeh99}
and they are shown to be extended from the western edge \citep{hewitt08}. A study of
surrounding material of the SNR using \spitzer\ IRAC images by
\citet{phillips09,phillips10} provides circumstantial evidence of interacting with
molecular clouds, but its direct evidence still lacks. There is no previous detection
of this SNR in optical or infrared. The distance to G357.7+0.3 is 6.4 kpc which is
consistent with -35 \kms\ (from OH lines) in Galactic rotation curve
\citep[see][]{yusef-zadeh99}.

In this paper, we report direct evidence that the SNR G357.7+0.3 interacts
with molecular clouds resulting in broad millimeter lines (such as CO and
HCO$^+$). The dynamic motion is revealed in the shocked clouds. We also
report detection of shocked H$_2$ emission in mid-infrared from G357.7+0.3
using \spitzer\ spectroscopy IRS data. Section 2 describes observations
which include 10 independent observing runs (see Table \ref{Tobs}). In
Section 3.1, we report direct evidence that the SNR G357.7+0.3 interacts
with molecular clouds with broad millimeter lines (such as CO and HCO$^+$)
using ground-based telescopes. Spectral mapping of CO lines is described in
Section 3.2 and 3.3. The detection of an atomic line of [C~II] at 158$\mu$m
and the upper limit of high-J CO line in Section 3.4. Large-scale molecular
cloud maps surrounding the SNR are presented in Section 3.5. The detection
of molecular hydrogen line with {\it Spitzer} IRS is in Section 3.6. In
Section 4, we discuss the physical conditions of shocked gas including
excitation of molecular hydrogen and CO gas. Our paper presents the first
direct evidence that the SNR G357.7+0.3 interacts with molecular clouds.
Because SNR-MC interactions are valuable astrophysical laboratories, our
paper provides a new and exciting laboratory where molecular
astro-chemistry and shocks can be studied.

\begin{table*}
\caption[]{Summary of Millimeter and Infrared Observations }\label{Tobs}
\begin{center}
\begin{tabular}{lllll}
\\
\hline \hline
Date  & Telescope & Lines \\
2003 May 18, 19 & HHSMT & CO(2-1)  \\
2003 Jun 8  & HHSMT & CO(2-1) \\
2005 August 14 & {Spitzer} IRS & H$_2$ \\
2006 March 3, 4 & HHSMT & CO(3-2) \\
2006 March 6-9 & ARO 12-Meter & $^{13}$CO(1-0), HCO$^+$   \\
2006 April 14, 15 & ARO 12-Meter & $^{13}$CO(1-0) maps \\
2007 September 19 &MOPRA     & HCO$^+$(1-0), HCN(1-0) etc \\
2013 July 18   & SOFIA GREAT & [C~II], CO(11-10) \\
2015 April 16 & APEX (Obs. 1) SHeFI  &  CO(2-1), $^{13}$CO(2-1)  \\
2015 August 16, 17 & APEX (Obs. 2)  FLASH, SHeFI  & CO(4-3), CO(3-2), $^{13}$CO(3-2)  \\
\hline \hline
\end{tabular}
\end{center}
\renewcommand{\baselinestretch}{0.8}
\end{table*}

\section{Observations}
We performed ten independent observing runs using  ground-based telescopes, {\it
Spitzer} Space telescope, and the Stratospheric Observatory
for Infrared Astronomy (SOFIA) air-borne telescope over a time span of
more than 10 years. The ground-based observing runs are using Heinrich Hertz Submillimeter
Telescope (HHSMT), 12-Meter (12-m, hereafter) Kitt Peak telescope, and Atacama
Pathfinder Experiment (APEX) telescope. The observational dates are summarized in
Table \ref{Tobs}. 
The positions of G357.7+0.3 we observed are listed
in Table \ref{Tg357pos} and marked on Figure \ref{g357radio}.
All maps in this paper have equatorial coordinates in J2000.

\begin{table}
\caption[]{Observed positions of G357.7+0.3 }\label{Tg357pos}
\begin{center}
\begin{tabular}{lllll}
\\
\hline \hline
Position & Offset &  R.A., Decl.\\
OH-A1/BML-A &(0,0)      &17:38:30.39,-30:33:17.7\\
APEX Obs. 2 & (0,0)      &17:38:30.39,-30:33:17.7\\
OH-A2&(+9.8,-23)      &17:38:30.39,-30:33:17.7\\
\hline
BML-B& (-60,-60)  &17:38:26.9,-30:34:17.7\\
{\it Spitzer} IRS  & (-60,-60)      & 17:38:26.9,-30:34:17.7\\
MOPRA     &  (-50,-50)           &17:38:27,-30:34:08 \\
SOFIA (H$_2$ peak)   & (-48,-48)    &  17:38:27.25,-30:34:06.4 \\ 
APEX Obs. 1 & (-60,-60)  &17:38:26.9,-30:34:08\\ 
\hline
BML-C &(-30,-90)  &17:38:28.7,-30:34:47.7\\
BML-D & (+60,+30) &17:38:38.8,-30:32:47.7\\
\hline \hline
\end{tabular}
\end{center}
\renewcommand{\baselinestretch}{0.8}
\end{table}

\subsection{HHSMT observations}

We performed mm and submillimeter (submm) observations of G357.7+0.3 using the HHSMT (or SMT)
{\footnote[1]{http://www.as.arizona.edu/aro/}}
located at Mt. Graham,
Arizona centered on one of
the OH maser positions, OH-A1 of G357.7+0.3 (R.A.\ $17^{\rm h} 38^{\rm m}
30.39^{\rm s}$ and Dec.\ $-30^\circ$ 33$^{\prime} 17^{\prime \prime}$,
J2000) on 2003 May 18-19 and June 8 and 2006 March 3-4. We observed
$^{12}$CO(3-2) and $^{12}$CO(2-1) using acousto-optic spectrometer (AOS)
and filterbank. The AOS spectra (see Figure \ref{smt12mspec}) were measured
with a 2048-channel, 1 GHz total bandwidth AOS with an effective resolution
of 1 MHz. The observations were made with three facility SIS mixer
receivers placed at the Nasmyth focus. The beam efficiencies of
single-polarization receivers in the frequency bands 210 –- 275 GHz and 430
–- 480 GHz are 0.77 (an example of CO(2-1)) and 0.45, and a
dual-polarization receivers covering the frequency band 320 –- 375 GHz is
0.48 for CO (3-2). The telescope beam size is 34$''$ at 217 GHz, 22$''$ at
347 GHz, and 18$''$ at 434 GHz. One of challenges of observation is to find
a clear reference position since the SNR is located at the Galactic plane;
we tried 2-10 positions and verified the emission from the reference
positions were clear. The final reference position used is RA.\ $17^{\rm h}
35^{\rm m} 15.5^{\rm s}$ and Dec.\ $-30^\circ$10$^{\prime} 48^{\prime
\prime}$ (J2000).

We performed spectral GRID mapping (10x10 spectra) in CO(2-1) covering
4.5$'$$\times$5$'$ area with a spacing of 30$''$ centered on BML-B position using HHSMT.
Details are described in Section 3.2. 

\subsection{12-Meter Telescope observations}

The 12-Meter Telescope is located on Kitt Peak and is one of the Arizona
Radio Observatory Telescopes.
We observed the SNR G357.7+0.3 on 2006, March 6-16 and the lines we
observed are $^{13}$CO (1-0) and HCO$^+$ (1-0) (see Table \ref{TCOlines}).
We observed a few positions for HCO$^+$ lines
and a large $^{13}$CO(1-0) map of 30$'$$\times$35$'$
area covering the SNR G357.7+0.3 and its surroundings.

\subsection{{\it Spitzer} IRS observations}

We performed an IRS spectral mapping centered on the northwestern shell of
G357.7+0.3 (R.A.\ 17$^{\rm h} 38^{\rm m} 26.9^{\rm s}$ and Dec.\
$-30^\circ$34$^{\prime} 17.7^{\prime \prime}$, J2000; position $``$BML-B"
which is a peak of CO broad molecular line (BML) as a part
of {\it Spitzer} IRAC GTO time (PI: Giovanni Fazio). 

The short-low (SL: 8-15 $\mu$m) covered 75$''$$\times$60$''$ (supplement
data covered the same area in the south for SL2, and in the north for SL1),
and long-low (LL) covered 170$''$$\times$55$''$ (supplementary data covered
the same area to the west for SL2, and to the east for LL1). The Long Low
(LL: 15-40 $\mu$m) IRS data were taken on 2005 August 14 with 6 cycles of
30 sec exposure time; this yields a total exposure time of 360 sec for the
first and second staring positions.  The SL IRS observations were made with
3 cycles of 60 sec exposure time and one cycle covers 2 dither positions;
this yields a total exposure time of 360 sec per sky position. The spatial
resolution (=1.2$\times$$\lambda$/D sr where D is 85 cm for {\it Spitzer})
of H$_2$ image generated from the IRS data is approximately 2$''$, 3$''$,
3.6$''$, 5$''$ and 8.3$''$ at 6.9, 9.6, 12.2, 17, and 28$\mu$m, respectively.
The IRS spectra (AORkey of 21819136) were
processed using the S18.8 pipeline products and reduced
using CUBISM. The spectra are extracted for the bright part of H$_2$
emission with a rectangular region of 40$''$$\times$50$''$ centered on the
IRS position of R.A.\ $17^{\rm h} 38^{\rm m} 26.9^{\rm s}$ and Dec.\
$-30^\circ$34$^{\prime}17.7^{\prime \prime}$ (J2000) as shown in Figure
\ref{g357radio}. The region is the overlapped region among SL1, SL2, LL2
and LL1.

\subsection{MOPRA observations}

Observations of selected positions were obtained using the 22m Australia
Telescope
MOPRA{\footnote[2]{http://www.narrabri.atnf.csiro.au/mopra/obsinfo.html}}
antenna during September 2007. G357.7+0.3 was observed using the MOPS
spectrometer backend configured in zoom mode to simultaneously observe 16
windows within an 8.3 GHz bandwidth in both linear polarizations. Each
window has 137.5 MHz sampled over 4096 channels. Final spectra (see Figure
\ref{mopraspec}) are smoothed with a four-channel Gaussian function, giving
a velocity resolution of 0.44 km s$^{-1}$ at 90 GHz. For all sources we
observed C$^{34}$S (2-1) at 96.4 GHz, CH$_{3}$OH (2-1) at 95.9 GHz,
CH$_{3}$OH (8-7) at 95.1 GHz, N$_2$H$^+$ (1-0) at 93.2 GHz, $^{13}$CS (2-1) at
92.5 GHz, HNC (1-0) at 90.7 GHz, HCO$^+$(1-0) at 89.2 GHz, and HCN (1-0) at
88.7 GHz. HCN J=1--0 is a triplet with F=2--1,1--1,0--1 at 88.631847,
88.630416 and 88.633936 GHz, respectively. When fitting this line we fit
the F=2--1 line and assume the two other lines of the triplet are found at
a fixed velocity separation of +4.84, -7.08 \kms\ and fixed line strengths
relative to F=2--1 of 0.5, 0.25 for the F=1--1, F=1--0 transitions
respectively. The main beam efficiency, T$_{mb}$, is 0.4--0.49 in the band
\citep{ladd05}.

HCO+ and HCN lines are detected as shown in Figure \ref{mopraspec} (see
Section 3.2 for details), and the upper limit of C$^{34}$S(2-1), CH$_{3}$OH
(2-1), CH$_{3}$OH (8-7), N$_2$H+(1-0), $^{13}$CS (2-1), and HNC (1-0) is
0.05 K for each.

\subsection{APEX observations}


APEX observations were conducted on 2015 April 16 and on 2015 August 16 and 17. The
observing time of 2015 April 16 observations was granted (PI: Andersen) from ESO open
proposal call, and 2015 August time is through APEX instrument PI allocated time. We
used SHeFi, FLASH345, and FLASH460 (Heyminck et al. 2006).
 The beam efficiency of CO(3-2) and CO (4-3) are 0.73,
and 0.60, respectively, and the beam sizes are listed in Table \ref{TCOlines}. The data
were reduced in CLASS{\footnote[3]{See http://www.iram.fr/IRAMFR/GILDAS}} and the final
results were exported into IDL.

\begin{table*}
\caption[]{Summary of Molecular line Properties \label{TCOlines}}
\begin{center}
\begin{tabular}{llrllrrll}
\\
\hline \hline
Position &line & Frequency & Telescope &   V$_{lsr}$ &   FWHM & $\int Tdv$ &RMS & t$_{int}$ \\
     &           &(GHz) & [beam size ($''$)]  & (km s$^{-1}$)  & (km s$^{-1}$) & (K km s$^{-1}$) &(K) & (min) \\
\hline
OH-A1 & CO (2-1) & 230.5379 & HHSMT [30] & -35.46$\pm$0.17 &    17.23$\pm$0.58  &89.94$\pm$1.83 & 0.125 &22\\ 
OH-A1 & CO (2-1) & 230.5379 &APEX SHeFI [30] & -35.85$\pm$0.16& 18.37$\pm$0.53 & 90.08$\pm$1.99 &0.106 & 11 \\
OH-A1 & CO (3-2)  &345.7959 &HHSMT [22] & -35.86$\pm$0.07  &17.33$\pm$0.20 & 163.64$\pm$1.35 & 0.189 & 18\\
OH-A1 &CO (3-2)  &345.7959 &APEX FLASH [22]   &-35.18$\pm$0.06 & 17.26$\pm$0.18&90.09$\pm$0.64&0.068 & 21\\
OH-A1 & CO (4-3) &461.0407 &APEX FLASH [13] & -34.65$\pm$0.04  &  14.71$\pm$0.10 & 64.29$\pm$0.33 &0.134  & 43\\
OH-A1  &$^{13}$CO(1-0)  & 110.2013 & ARO 12-m [47] &-35.07$\pm$0.10 & 3.95$\pm$0.31 & 4.57$\pm$0.26& 0.100 &20\\
OH-A1  &$^{13}$CO(2-1) &220.3986 &APEX SHeFI (30) &-34.89$\pm$0.07 & 7.52$\pm$0.23 & 7.09$\pm$0.16 & 0.056 &27\\  
OH-A1 & $^{13}CO$(3-2) &330.5879 &   APEX FLASH [22]  &-34.73$\pm$0.14& 10.76$\pm$0.39 &4.22$\pm$0.12 & 0.040 &21 \\
OH-A1 &HCO$^+$ &89.1885 & ARO 12-m [58] & -33.47$\pm$0.34 & 25.03$\pm$0.94 & 4.04$\pm$0.12 & 0.017  &123 \\ 
\hline \hline
BML-B  &CO (2-1) & 230.5379 &APEX-1 (SHeFI) [28]&  -34.04$\pm$0.09& 26.79$\pm$0.19 &  111.52$\pm$0.71  &0.077& 6\\
BML-B  & CO (3-2) &345.7959      & HHSMT [22]  & -35.21$\pm$0.08  &  22.01$\pm$0.19  &  200.50$\pm$1.42&0.145 & 14 \\
BML-B & $^{13}CO$(2-1) &220.3986  & APEX (SHeFI) [28]  &    -34.932$\pm$0.231 & 16.67$\pm$0.60 & 6.28$\pm$0.18 &0.073 & 6\\
BML-B &HCO$^+$ (1-0)& 89.1885& ARO 12-m [58] &  -33.56$\pm$0.01 & 27.96$\pm$0.02 & 4.88$\pm$1.30 &0.012  & 98 \\
BML-B &HCO$^+$ (1-0)& 89.1885&MOPRA [38] &-34.75$\pm$0.01&25.16$\pm$0.01 &3.16$\pm$0.82 &0.023 &68\\
BML-B &HCN (1-0) &88.6316&MOPRA [38] &-32.16$\pm$0.01&27.25$\pm$0.01 &2.67$\pm$0.74 &0.024 & 68\\
BML-B & [C~II]   & 1900.5369 & SOFIA GREAT [14.1] & -30.32$\pm$1.54 & 15.69$\pm$3.34 & 5.61$\pm$1.08 & 0.110 &4.9\\
\hline \hline
\end{tabular}
\end{center}
\renewcommand{\baselinestretch}{0.8}
\end{table*}

\subsection{SOFIA Observations}

We observed the SNR G357.7+0.3 with the German Receiver for Astronomy at
Terahertz Frequencies (GREAT) \citep{heyminck12} on board the SOFIA
airborne observatory \citep{young12}. GREAT is a far-infrared high resolution
spectrometer with a resolving power of $\sim$10$^6$. Only low frequency
detectors covering 1.25 - 1.5 THz (wavelengths of 240 - 200$\mu$m; L1)
and 1.82 - 1.92 THz (165 - 156$\mu$m; L2) were available
during the cycle 1. The observation was a program (proposal ID of O1\_0059; PI:
Hewitt) of Guest Observer cycle 1 campaign. Five hrs of observation time
was awarded, but a total of $\sim$ 2 hrs ($\sim$1 hr on July 18 and $\sim$1
hr on July 28) were included in the flight series and the flights  for
remaining observing time of 3 hrs were cancelled. The observation of G357.7+0.3 took
place on 2013 July 18 toward the peak of H$_2$ emission as listed in Table
\ref{TCOlines}. 
The observed integration time for [C~II] and CO (11-10) lines are 5 min for
each, because  the first block of observations toward the first position of
G357.7+0.3 was lost due to tracking and wobbler issues. Successful
observing time was about 30 min on July 18. Note that the efficiencies of
SOFIA flights and observations have been significantly improved since 2014.
The observations made a map of 1$'$$\times$1$'$, but since the signal is
very weak, we have averaged the spectrum over the area (see
Figures \ref{greatciiline} and \ref{greatcoline}). The main beam
efficiencies of 0.67 were used for both of the GREAT channels L1 \& L2.

\begin{figure}
\psfig{figure=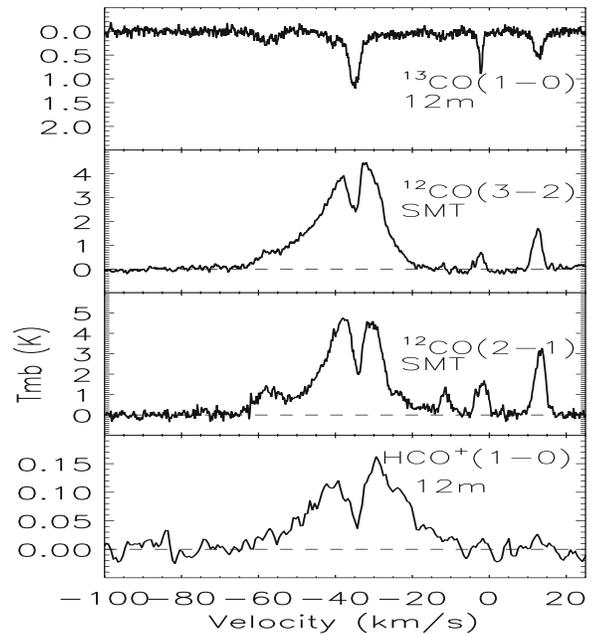,height=10truecm,width=9truecm,angle=90}
\caption{CO and HCO$^+$ spectra of G357.7+0.3 using HHSMT and 12-m telescope. The velocity range is from -100 to 25 \kms.
$^{13}$CO(1-0) line is shown from positive to negative, in order to demostrate the emission causes the absorption dip in 
the broad lines of $^{12}$CO(3-2), $^{12}$CO(2-1), and HCO$^{+}$.} 
\label{smt12mspec}
\end{figure}

\begin{figure}
\includegraphics[scale=0.4,angle=0]{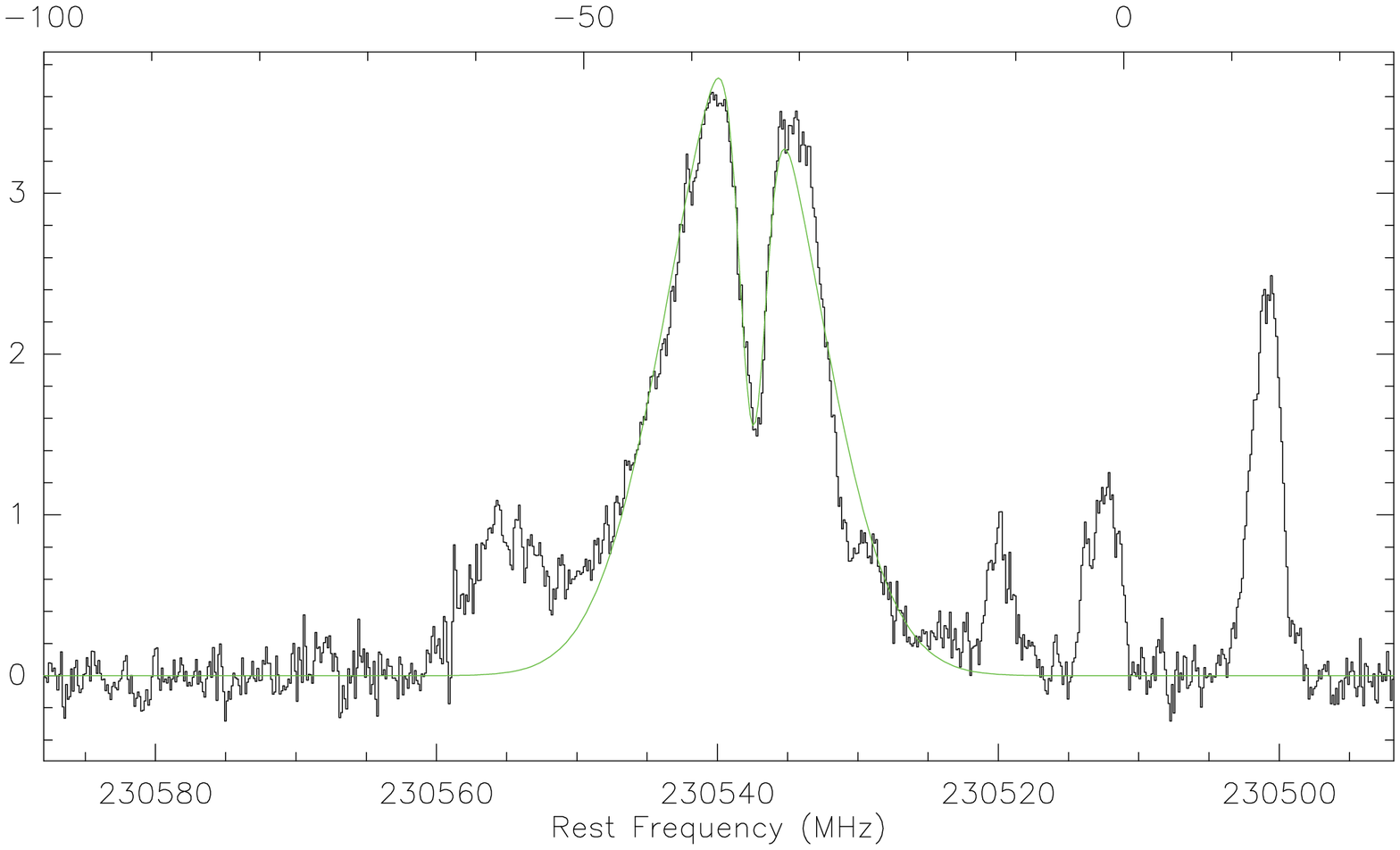}
\caption{HHSMT CO(2-1) spectrum superposed on spectral model fit. 
The fit is a combination of a broad emission line between -55 and -20 \kms\ and a narrow absorption 
line between -32 and -38 \kms.}

\label{specfit}
\end{figure}

\begin{figure}
\psfig{figure=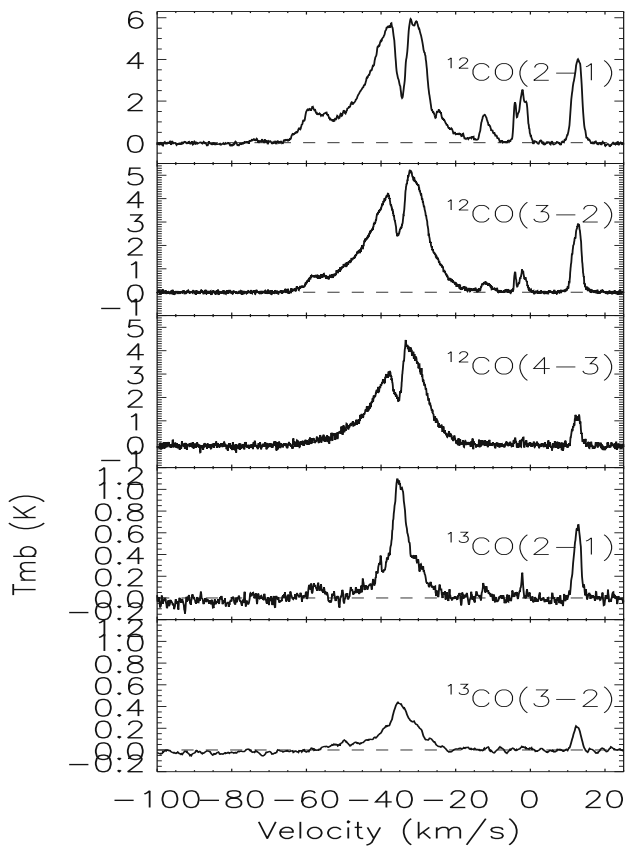,height=9.7truecm,width=12truecm,angle=90}
\caption{APEX spectra of G357.7+0.3 in CO(2-1), CO(3-2), CO(4-3), $^{13}$CO (2-1), and  $^{13}$CO (3-2) lines.}
\label{apexspec}
\end{figure}

\section{Results}
\subsection{Broad millimeter lines from shocked gas}

\begin{figure*}
\includegraphics[scale=1.,angle=90,width=15truecm,height=21truecm]{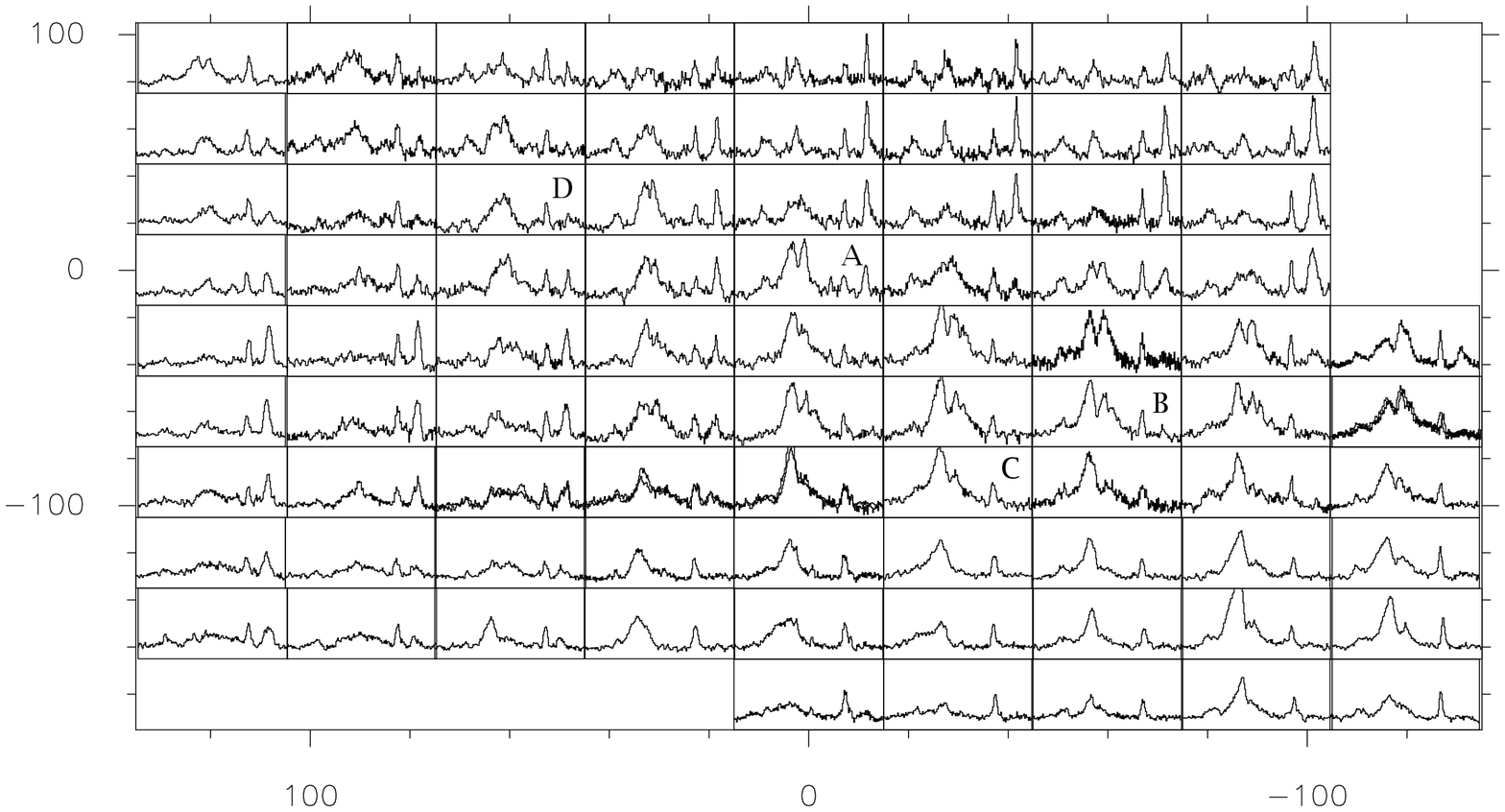}
\caption{GRID spectra of CO (2-1) of G357.7+0.3 with a 
spacing of 30$''$ (units of x- and y-axis
are offsets ($''$) from the OH-A1 position).
The BML-A, BML-B, BML-C and BML-D positions (see Table \ref{Tg357pos}) with representative CO spectra 
are marked as A, B, C, and D, 
respectively.
Broad lines appear in extended regions, and the structure shows elongated from northeast to
southwest. The individual spectra have X-axis in velocity from -100 to 25 \kms\ in velocity, and y-axis
in temperature from -0.5 to 5.8 in T$_{mb}$(K).
}
\label{cogridspectra}
\end{figure*}

\begin{figure*}
\epsscale{0.85}
\includegraphics[scale=0.64,angle=0]{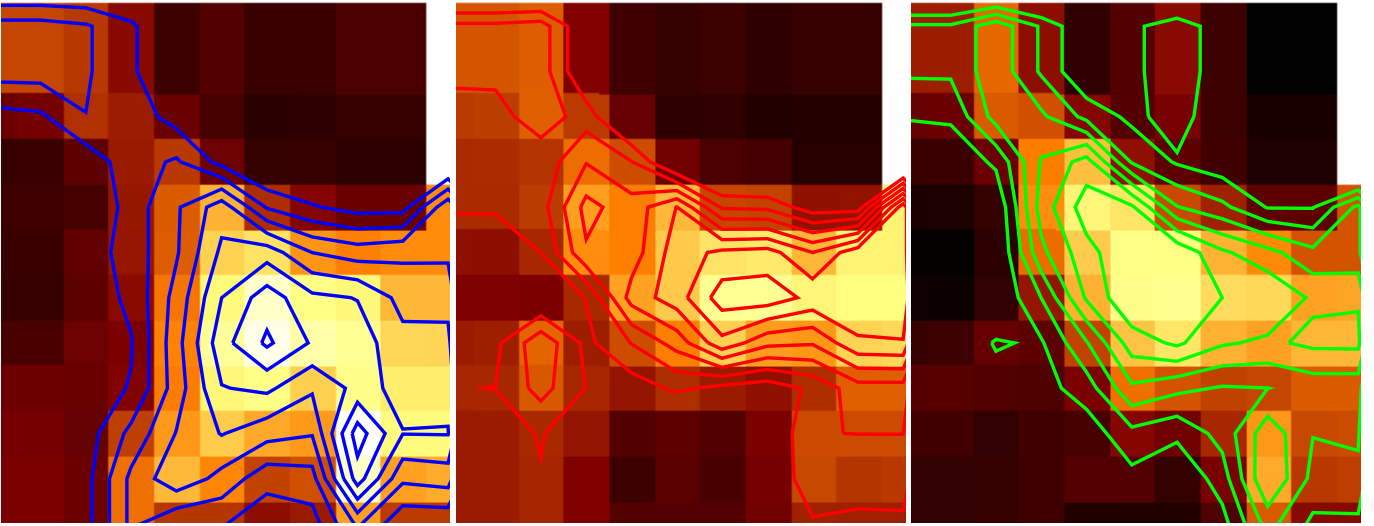}
\caption{CO(2-1) broad line maps in three velocity ranges (R.A. and Decl. are
labeled in Figure \ref{cowing2}). Image of blue wing (left in blue
contours) is from the velocity ranges between -53 and -38 \kms, red wing
image (middle; in red contours) between -31 and -27 \kms, middle velocity
image where the spectrum shows broad line with self-absorption (right; in
green contours is between -38 and -31 \kms.
Contours on blue wing image (left) are 13, 20.8, 28.7, 36.6, 44.4, 52.3, 20.14, and 68 K
\kms, on red wing image (middle) are 8, 10.5, 13.1, 15.7, 18.3, 20.8, 23.4, 23.4 and 26 K
\kms,  and on the middle velocity map are 14, 17.8, 21.7, 25.5, 29.3, 33.3 and 37.0 K
\kms.}
\label{cowing1}
\end{figure*}

\begin{figure*}
\plottwo{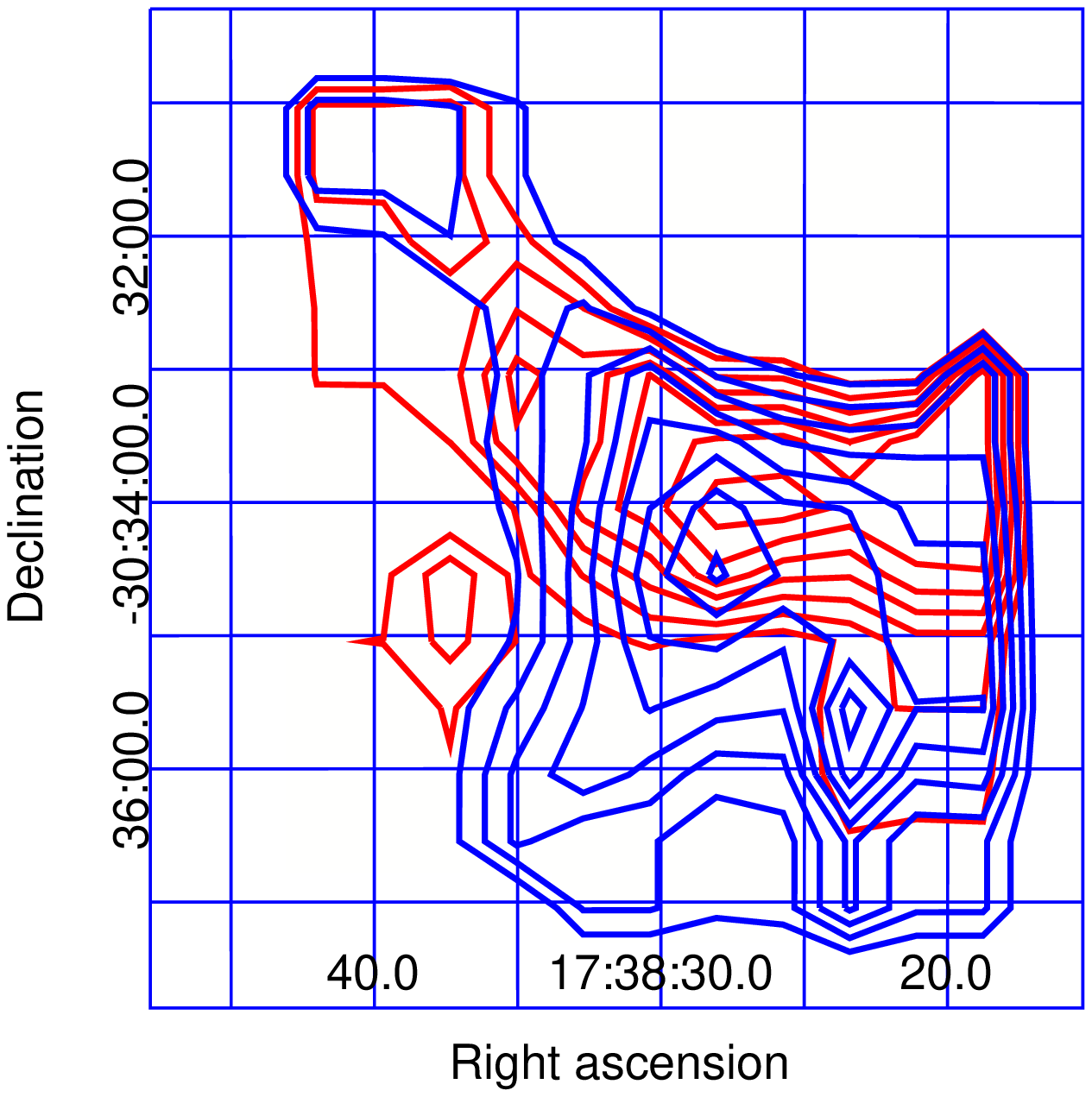}{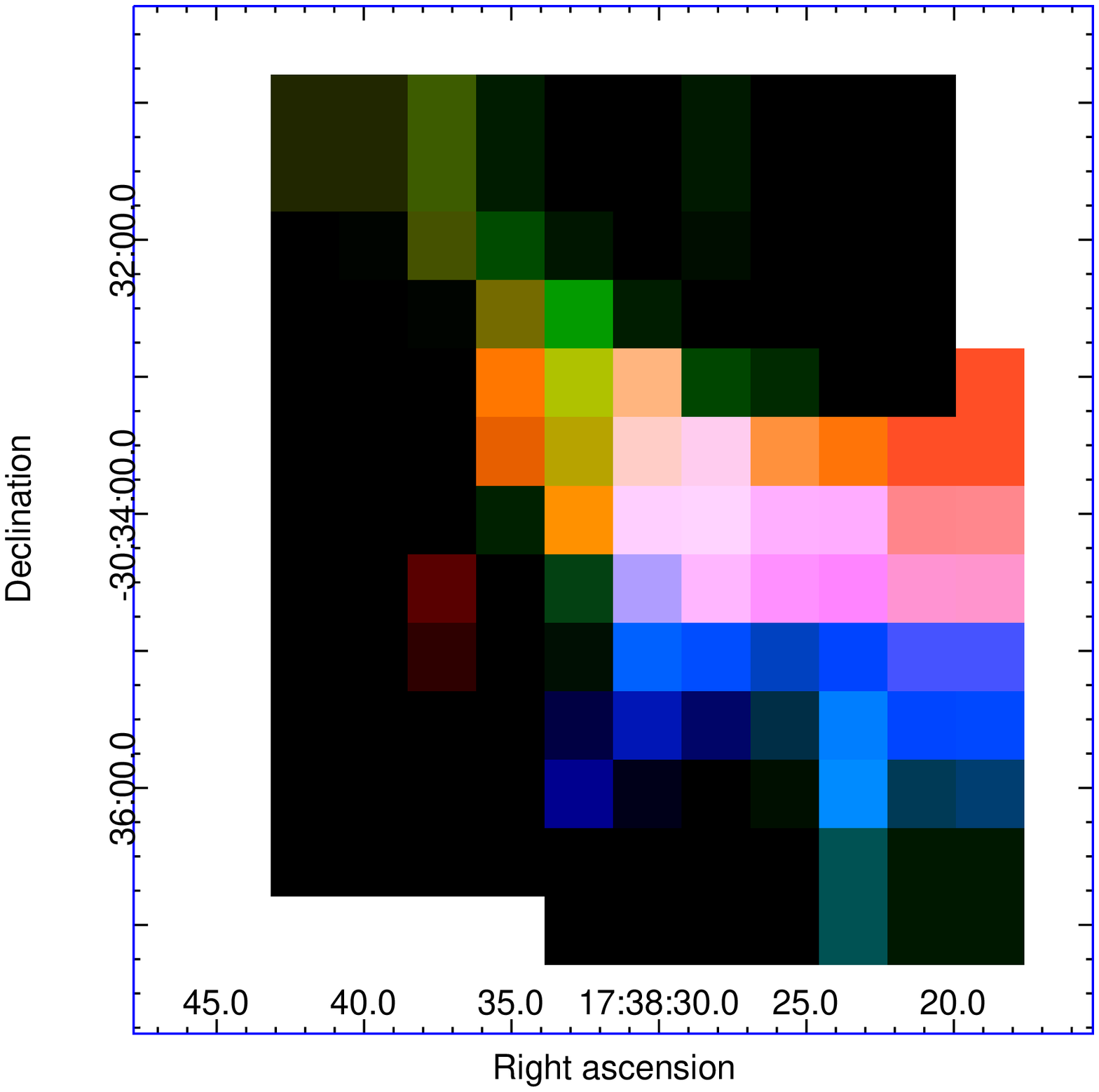}
\caption{Comparison of blue and red wing images from CO(2-1) shown in Figure
\ref{cowing1} (left). Three color images of three velocity range maps from
the broad line shown in Figure \ref{cowing1}; blue wing, red wing and
middle velocity map are represented in blue, red and green, respectively
(right).}
\label{cowing2}
\end{figure*}

Figure \ref{smt12mspec} (the second panel) shows the $^{12}$CO (3-2) line
toward the OH-A1 position of G357.7+0.3 observed with HHSMT. 
The spectrum
shows a broad line  with a FWHM of 17.32 \kms\ at a velocity of -35.86 \kms
(see Table \ref{TCOlines}) and a narrow ($\sim$ 4\kms) absorption dip at
-35.30 \kms. Such a broad line is caused by strong SN shocks passing
through dense molecular clouds, and the detection of such broad lines is
direct, unambiguous evidence that G357.7+0.3 is a SNR interacting with
molecular clouds. Figure \ref{g357radio} shows the area
(over 4.5$'$$\times$5$'$) where broad lines are detected (see Section 3.2
for details). Figure \ref{smt12mspec} includes other millimeter molecular
lines of $^{13}$CO(1-0), $^{12}$CO(2-1), and HCO$^+$ observed with HHSMT
and 12-m telescopes. HCO$^+$ spectrum has a lower signal-to-noise than
those of CO(3-2) and CO(2-1), but the line profile is the same as those of
the other two lines, indicating that the HCO$^+$ line also shows a broad
line with a self-absorption.  Both of $^{12}$CO(2-1) and HCO$^+$ spectra
show similar profiles to that of $^{12}$CO(3-2) with the widths of
$\sim$20-30 \kms. We fit each spectrum either using a gaussian profile with
a mask of the velocity range of the self-absorption line, or two gaussian
components with a broad emission line and a narrow absorption line.
The results from the two methods yielded similar results and are summarized in Table \ref{TCOlines}.  
The RMS noises are obtained using a long baseline typically between -200 and 100 \kms.
An example of a spectral fit is shown in Figure \ref{specfit}.
The line properties (e.g. for broad lines) are summarized in Table \ref{TCOlines}.

We observed additional CO lines with  APEX, which offers  simultaneous
observations of three lines. The spectra of $^{12}$CO(2-1), $^{12}$CO(3-2),
$^{12}$CO(4-3), $^{13}$CO(2-1) and  $^{13}$CO(3-2) are shown in Figure
\ref{apexspec}. The CO lines of $^{12}$CO(4-3), $^{13}$CO(2-1) and
$^{13}$CO(3-2) were only observed with APEX. The $^{13}$CO(3-2) line shows a broad
line, and $^{13}$CO(2-1) show a combination of a broad line similar to the one in
$^{13}$CO(3-2) and a narrow emission line similar to the one in $^{13}$CO(1-0).

\subsection{Spectral mapping of CO(2-1)}

Spectral GRID mapping was made in CO(2-1) using the HHSMT for a
4.5$'$$\times$5$'$ area. The GRID spectra are shown in Figure
\ref{cogridspectra}, and a typical RMS of CO(2-1) line is given in Table
\ref{TCOlines}. The broad line structures are extended over a
4.5$'$$\times$5$'$ region (the entire GRID map we observed) and elongated
from northeast to southwest. Three maps of the blue wing (with a velocity
range between -58 and -38 \kms), red wing (between -31 and -27 \kms), and
the middle velocity, the broad line with the self-absorption (between -38
and -31 \kms) are shown in Figures \ref{cowing1} and \ref{cowing2}.
We chose these velocity ranges based on the CO(2-1)
spectra in order to avoid materials that are not related to the shocked gas
in the SNR G357.7+0.3. The CO spectra show two weak lines at the velocity
-56 \kms\ and at -23 \kms\ (between -27 to 20 \kms) on the top of the broad
lines (see Figures \ref{smt12mspec} and \ref{apexspec}) and the materials
at these velocities are likely unrelated to the SNR. These features are
less noticeable at higher transition lines of CO(4-3) and CO(3-2). The
blue wing, the material moving toward us, is located in the south relative
to the pre-shock gas (in green; this emission may include pre-shock gas)
and the red wing, the material moving away from us, is located in the north
as shown in Figure \ref{cowing2}.

\begin{figure*}
\plotone{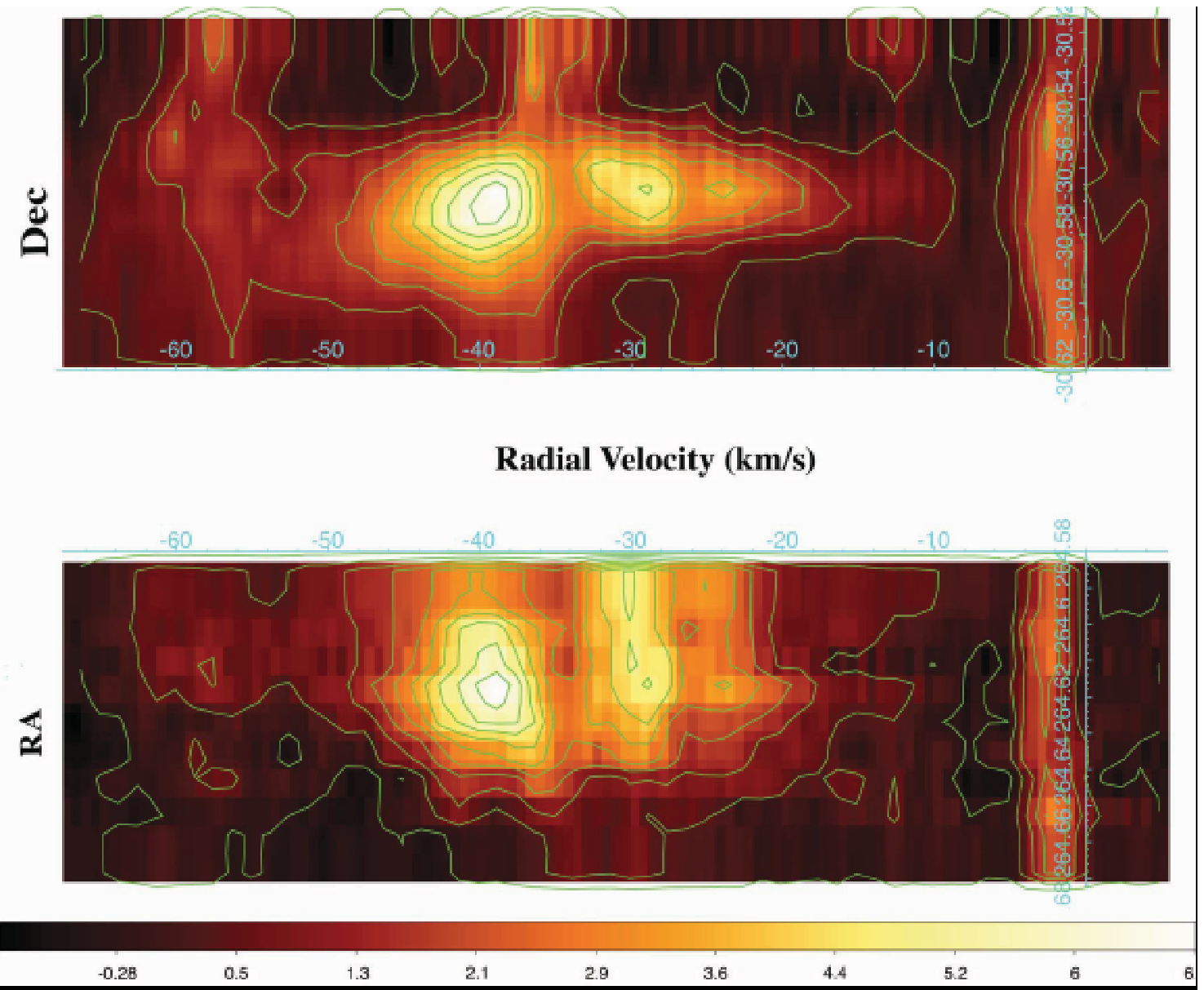}
\caption{CO(2--1) position-velocity maps (a: top) cut at constant R.A. = $17^{\rm h}
38^{\rm m} 26.9^{\rm s}$ and (b: bottom) cut at constant decl=$-30^\circ$
33$^{\prime} 57.07^{\prime \prime}$. Broad lines from -50 \kms\ to -10 \kms\
with an absorption dip at 34 \kms.  The broad line feature continues outside the
map in R.A. direction. Contours are 0.36, 1.07,1.78, 2.49, 3.20, 3.91, 4.63,
5.45, 6.05 and 6.76 K. T he positions are labeled in decimal degrees, for both
(a) declination and (b) right ascension.}
\label{positionvelmap}
\end{figure*}

Figure \ref{positionvelmap} shows position-velocity maps that slice through
the spectral data centered on BML-B (the center of IRS map) in R.A. and
Decl direction, respectively. The exceptionally wide velocity dispersion of
the broad molecular regions compared to the ambient gas (which has a
typical width of $<$7 \kms\ FWHM; for example at $\sim$0 \kms) is evident.
The map shows broad lines from -50 \kms\ to -10 \kms\ with an absorption
dip at 34 \kms. The broad line feature continues outside the map in R.A.
direction.

\begin{figure*}
\includegraphics[scale=1.2,angle=90,width=18truecm,height=11truecm]{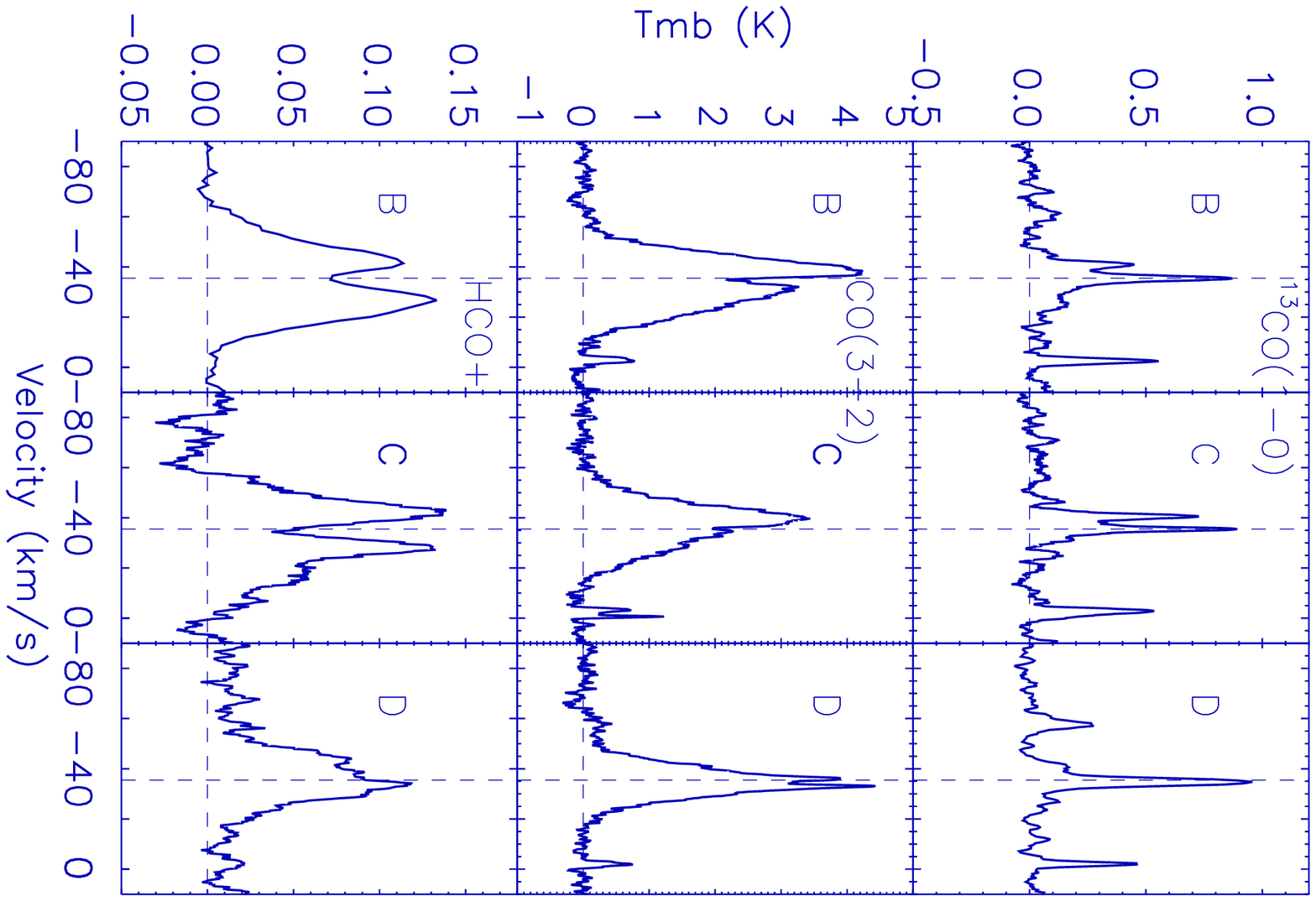}
\caption{Comparison of $^{13}$CO(1-0), CO(3-2) and HCO$^+$ spectra for the
positions of BML-B,  BML-C and BML-D. CO (3-2) data are taken with HHSMT,
and  $^{13}$CO(1-0) and HCO$^+$ data are taken with 12-m telescope.}
\label{g357bcdspec}
\end{figure*}

\begin{figure}
\includegraphics[scale=0.25,angle=90,height=5.5truecm,width=7truecm]{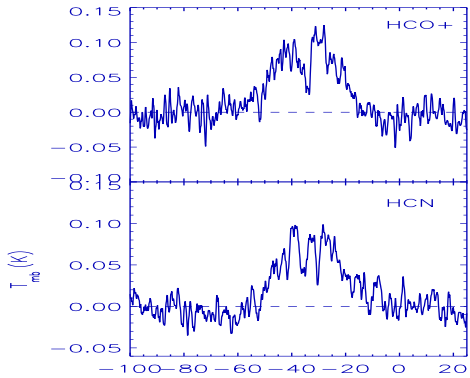}
\caption{MOPRA spectra of HCO$^+$(1-0) and HCN(1-0) toward the position of BML-B.}
\label{mopraspec}
\end{figure}

\begin{figure}
\includegraphics[scale=0.25,angle=90,height=6.5truecm,width=8.5truecm]{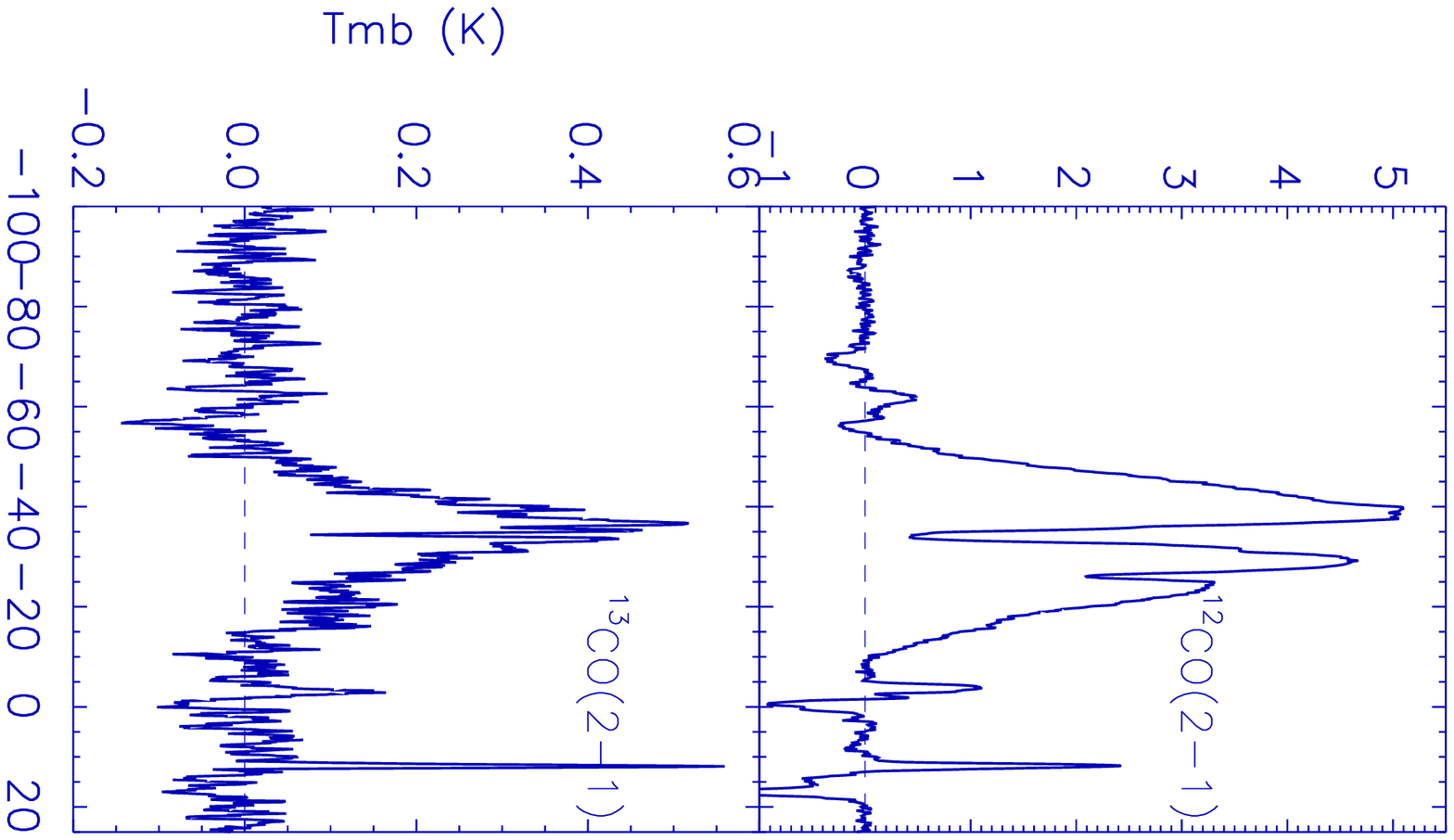}
\caption{APEX spectra of $^{12}$CO(2-1) and $^{13}$CO(2-1) toward the position BML-B. }
\label{apexspec2}
\end{figure}

We named three representative broad molecular line (BML) positions from the
CO(2-1) grid spectra as BML-B, BML-C and BML-D in addition to the OH-A1
position (we call this position BML-A). The positions are
marked in Figure \ref{g357radio} and listed in Table \ref{Tg357pos}. 
BML-B is the peak of broad lines and has broader CO lines than the
OH-A1(BML-A) and other positions. We note that the peak of the CO broad
lines does not coincide with any of the OH positions as shown in Figure
\ref{g357radio}.
We made ARO 12-m, {\it Spitzer} IRS, MOPRA, SOFIA, and additional APEX
observations toward the BML-B position or its vicinity. The spectra of
$^{13}$CO(1-0), CO(3-2) and HCO$^+$ toward BML-B, BML-C and BML-D are shown
in Figure \ref{g357bcdspec}. CO(3-2) and HCO$^+$ spectra show broad lines
with the widths of up to 27 \kms\ from shocked gas (see Table
\ref{TCOlines}) and $^{13}$CO(1-0) shows narrow components from cold gas.
The HCO$^+$ taken with the 12-m telescope shows the broadest line with a
FWHM of 27.96$\pm$0.02 \kms\ we detected as shown in Figure
\ref{g357bcdspec}. The MOPRA spectrum also detects a broad line of HCO$^+$
as shown in Figure \ref{mopraspec}, although the signial-to-noise is not as
good as that from the 12-m telescope. As we have seen in the position of
OH-A1, the broad lines of CO or HCO$^+$ spectra show anti-correlation with
$^{13}$CO(1-0). APEX spectra toward BML-B in Figure \ref{apexspec2} show
detections of CO(2-1) and $^{13}$CO(2-1). The lines are broad with widths
of 26 and 16 \kms, respectively (see Table \ref{TCOlines}). The line
profiles also show self-absorption lines, but the $^{13}$CO(2-1) line shows
only a very small amount of self-absorption. MOPRA additionally detects a
broad line of HCN with a FWHM of $\sim$25 \kms.

\subsection{Self-absorption line in broad molecular line}

CO or HCO$^+$ spectra toward BML-B, BML-C, and BML-D show similar
anti-correlation of  the broad lines of CO(3-2) and HCO$^+$ with $^{13}$CO(1-0)
as shown in Figure \ref{g357bcdspec}. In Table \ref{Tg357pos}, the positions of
B, C, and D are where CO(2-1) spectra reveal representative BMLs.

The narrow line components are unshocked clouds which are likely
a part of parent molecular clouds of the shocked CO gas because they are
approximately at the velocity similar to that of shocked gas. The line-of-sight
absorption is stronger for the CO(2-1) transition, which can be absorbed by cold
gas with substantial populations in the level J=1, than for the CO(3-2) or
CO(4-3) transitions. In other words, the $^{13}$CO(1-0) emission, which traces
the total column density and is dominated by the cold gas along the line of
sight, matches very well the center and width of the apparent ``notch'' cut out
of the $^{12}$CO spectra. 

The optical depth of CO emission from low-lying energy levels of gas in quiescent
molecular clouds is generally greater than unity.  
For HCO$^+$(1--0) the transition is out of the ground state, so even cold gas
is readily detected in absorption. The precise correspondences between the $^{13}$CO
emission and the apparent $^{12}$CO and HCO$^+$ absorption notches indicate they are
due to cold molecular gas in the parent molecular cloud. Conversely, the {\it lack}
of $^{13}$CO emission from the broader component that is so bright in $^{12}$CO
indicates it is optically {\it thin} (see Section 4.3 for detailed discussion) and 
due to a smaller column density of
more-highly-excited gas. A similar combination of broad emission  with narrow
superposed absorption lines was observed in the molecular-cloud-interacting SNRs W44
and W28 \citep{reach05}. When there is bright emission from hot, shocked gas behind
cold, unshocked gas, an absorption component appears. The $^{12}$CO (2-1) line also has 
narrow components at -57.9, -13, -1.8 and 13 \kms; these narrow components (line
widths $<$4 \kms)  are from cold gas in other molecular clouds along the line of
sight, unrelated to G357.7+0.3.

The narrow absorption line in the spectra of $^{12}$CO(4-3),  $^{12}$CO(3-2),
$^{12}$CO(2-1) and HCO$^+$ is anti-correlated with $^{13}$CO(1-0) where the narrow
line appears in emission (see Figures \ref{smt12mspec} and \ref{apexspec}).
The $^{13}$CO(3-2) line in Figure \ref{apexspec} shows a broad wing which is
from shocked gas, and $^{13}$CO(1-0) shows a narrow line with a line width of
3.9\kms which is pre-shocked gas. The $^{13}$CO(2-1) shows a combination of the
two, a broad and narrow components. One gaussian component fit to $^{13}$CO(2-1)
yielded a line width of 7.5 \kms as listed in Table \ref{TCOlines}. When we fit
with two gaussian components, the fit yielded a broad component with a width of
8.57$\pm$0.43 \kms and a narrow component with a width of 2.22$\pm$0.14
\kms. This is consistent with the idea that $^{13}$CO(2-1) shows a combination
of pre-shock and post-shock CO emission.

\subsection{SOFIA GREAT spectra}
\begin{figure} 
\includegraphics[scale=0.4,angle=0]{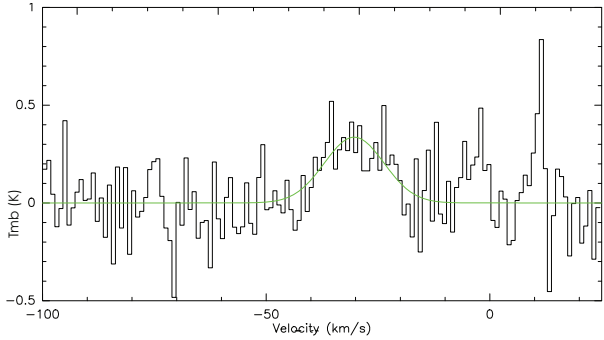}
\caption{SOFIA GREAT spectrum of [C~II] line is superposed on a gaussian fit. The width 
of the line is $\Delta$V=16 \kms\ and the line
 profile of [C~II] is similar to those of CO lines.
 }
\label{greatciiline}
\end{figure}

\begin{figure} 
\includegraphics[scale=0.37,angle=270]{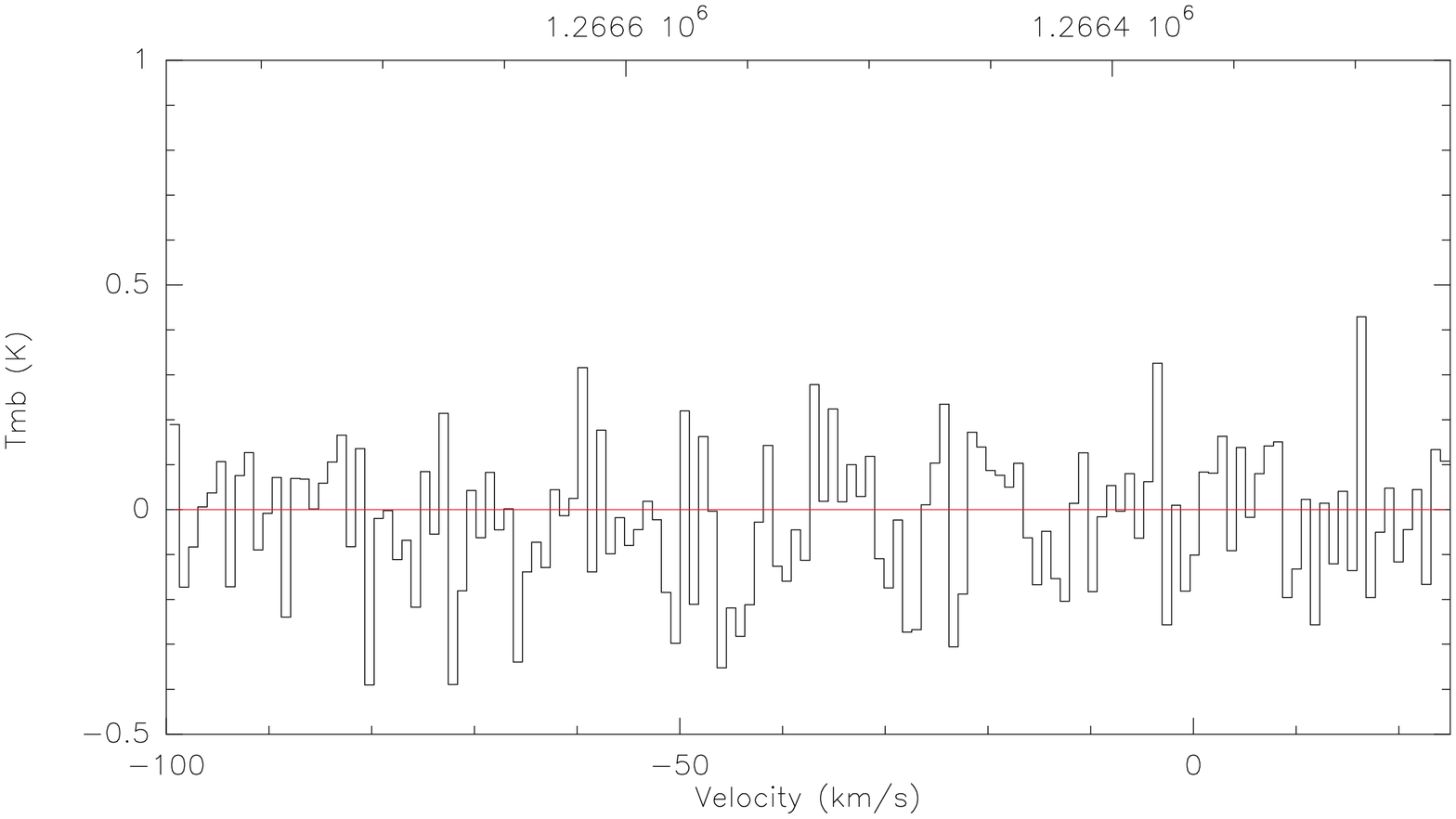}
\caption{SOFIA GREAT spectrum of CO(11-10) line.}
\label{greatcoline}
\end{figure}

In Figure \ref{greatciiline} the SOFIA GREAT spectrum of [C~II] at
158$\mu$m shows a broad line with a FWHM of 15.7 \kms\ (see Table
\ref{TCOlines}). Although the detection is only 3$\sigma$, the line profile
and the FWHM of [C~II] is similar to those of broad CO lines. The line
strength is equivalent to 2.84$\times$10$^{-13}$ erg~s$^{-1}$~cm$^{-2}$.
Using a beam size of 14.1$''$, the surface brightness is
7.85$\times$10$^{-5}$ erg~s$^{-1}$~cm$^{-2}$ sr$^{-1}$. Ionic lines
commonly originate from high velocity J-shocks \citep{hollenbach89, rho01,
reach00, hewitt09}. A resolved [O~I] spectrum of another SNR 3C391 at
63$\mu$m using ISO LWS shows a line width of 100\kms. This indicates that
atomic fine-structure lines such as [O~I]  and [C~II] oxygen can originate
from J-shocks. However, the line width of [C~II] in G357.7+0.3  is small,
15.7 \kms,  similar to the broad component of CO lines. This suggests that
[C~II]  comes from the same shock responsible for the CO lines. In other
words,  [C~II] is from a low ($\sim$15 \kms) velocity shock.
\citet{draine83} show that ionic lines could be more important coolants
than molecular lines even for low velocity ($<$20 \kms) shock \citep[see
Figure 4 of][]{draine83}. Unfortunately the models do not include [C~II] at
158$\mu$m itself, so we cannot directly compare the line brightness of
[C~II] with the model. Non-quasi-steady model of CJ-shocks (introducing a
J-type discontinuity in a C-type flow at a point in the steady-state
profile that is located in downstream of the shock) show enhanced cooling
by ions (e.g. oxygen) before molecular line cooling such as H$_2$ and CO as
described by \citet{lesaffre04a}. These spectra may be the first direct
evidence of such CJ shocks. The [C~II] line is known to be an important
cooling line in J-shock \citep{hollenbach89} and the detection of [C~II]
with the width of 16 \kms\ indicates that cooling by [C~II] may be moderate
in C-shock or CJ-shock in addition to cooling by various molecular lines
\citep{kaufman96}. The SOFIA spectrum of CO (11-10) line is not detected as
shown in Figure \ref{greatcoline} and its upper limit is 0.14 K which is a RMS noise estimated 
between -200 to 100 \kms.

\subsection{Molecular Cloud Maps Surrounding the SNR} 

\begin{figure}
\includegraphics[scale=1,angle=0,width=10.5truecm]{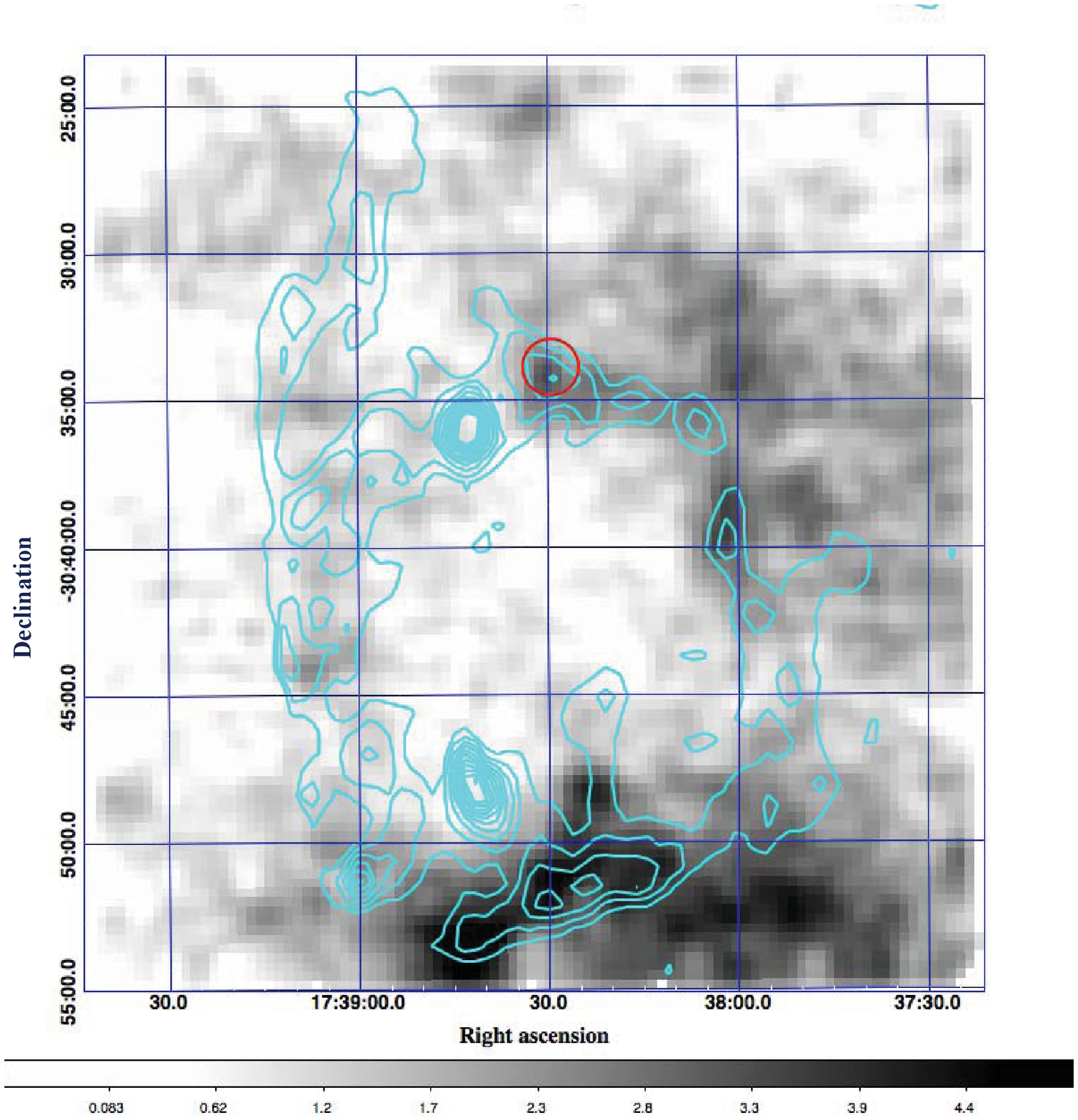}
\caption{$^{13}$CO(1-0) image of G357.7+0.3 integrated over velocities -41 to -31
\kms (greyscale) obtained with the ARO 12-m telescope. The greyscale bar
indicates the summed brightness in K; to convert to K~km~s$^{-1}$,
multiplied by 2.72 \kms. Overlaid on the $^{13}$CO image are cyan-colored
contours of the non-thermal radio brightness from the MOST Galactic Centre
Survey \citep{gray94}. The red circle indicates the location of the OH-A1
1720 MHz maser, where broad molecular lines are seen. The image is centered
on R.A.\ $17^{\rm h} 38^{\rm m} 31.9^{\rm s}$ and Decl.\
$-30^\circ$39$^{\prime} 19^{\prime \prime}$ (J2000) with a FOV of
$29.4'\times 30.1'$. }
\label{12m13comap}
\end{figure}

\normalsize

\begin{figure*}
\plotone{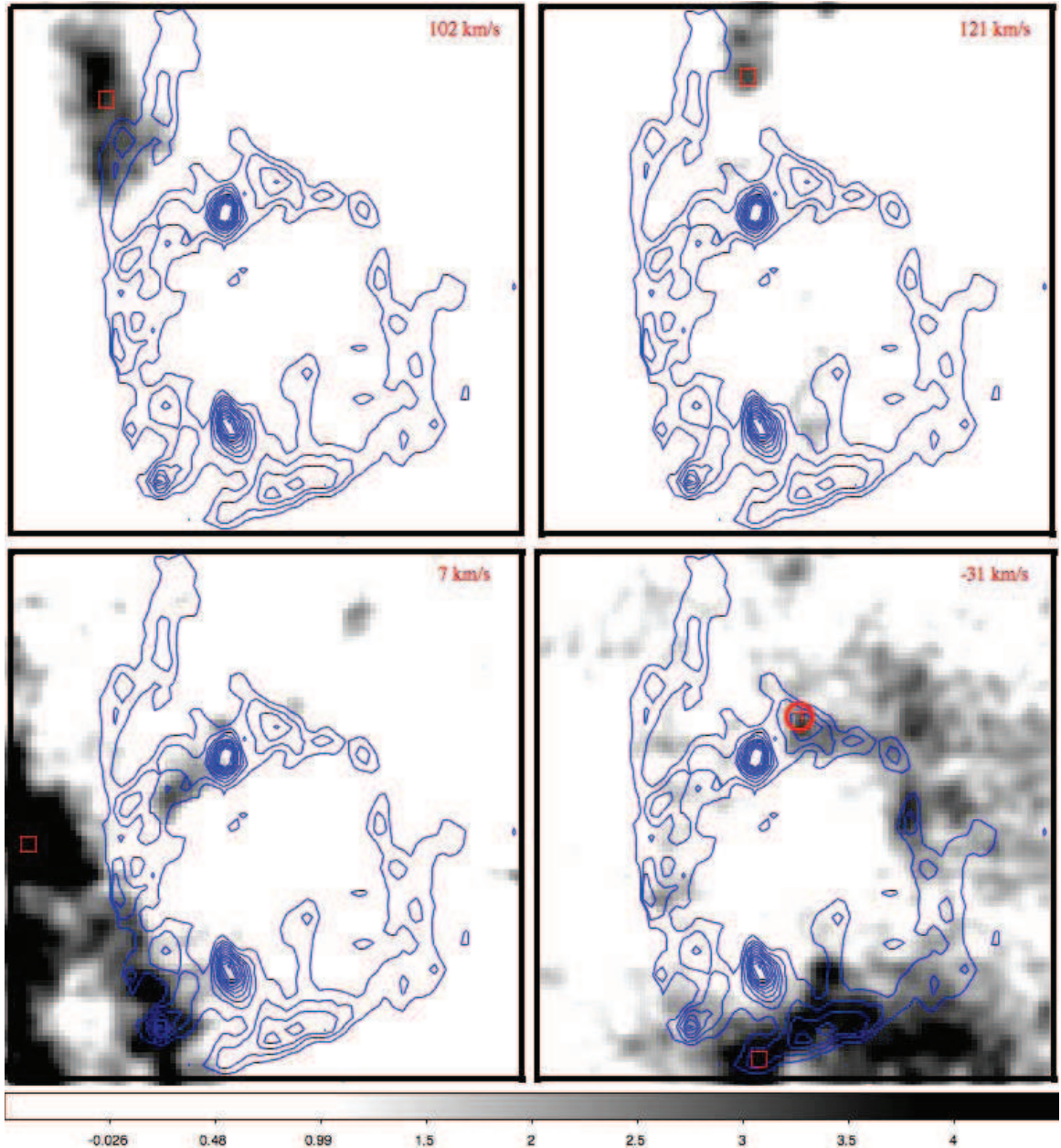}
\caption{Maps of $^{13}$CO (1-0) emission toward the SNR G357.7+0.3 integrated over
4 velocity ranges. Each panel is labeled in the top right with its velocity
in red, with the location of the highest intensity in each panel as a red
square. The blue contours repeated on each panel  are the non-thermal radio
brightness. The area showing the broad $^{12}$CO lines toward the SNR is
marked as a circle on the $^{13}$CO (1-0) map with -31 \kms. The scale bar
below the images for the greyscale $^{13}$CO images is in units of  antenna
temperature summed over channels; to convert to brightness temperature
integrated over velocity (K~km~s$^{-1}$), multiply by 2.3.} 
\label{12m13cochannelmap}
\end{figure*}

We have mapped a large-scale molecular cloud structure surrounding the SNR
G357.7+0.3 using $^{13}$CO(1-0) line that typically traces cold clouds. The
spatial resolution of the map is 47$''$ and the map covers
32$'$$\times$32$'$ larger than the size of SNR 24$'$$\times$24$'$.
$^{13}$CO(1-0) image of G357.7+0.3 integrated over velocities -41 to -31
km~s$^{-1}$ obtained with the ARO 12-m telescope is shown in Figure
\ref{12m13comap}. The supernova appears to be bounded by molecular gas, in
particular its southern, western, and northwestern portions. The sharpness
of the eastern radio continuum boundary in these contours is exaggerated
because of taper of the MOST field of view.

\begin{figure}
\includegraphics[width=8cm]{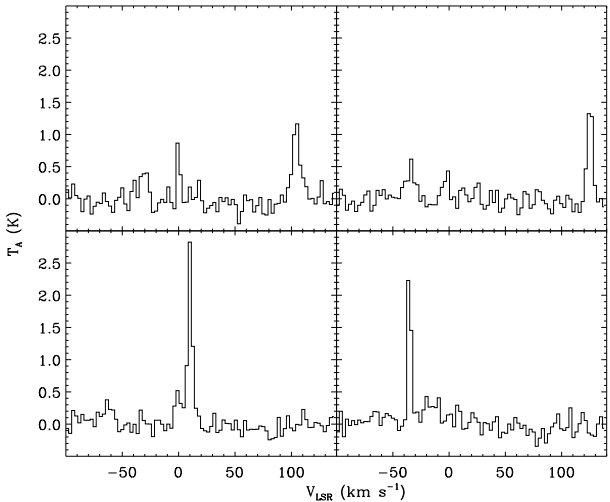}
\caption{Respresentative $^{13}$CO(1-0) spectra of 4 velocity channel maps in Figure
\ref{12m13cochannelmap}. The extracted regions are marked as boxes located
in the northeast on the +102 \kms\ velocity map, in the north on the +121
\kms, in the east on the 7 \kms\ map, and in the southern region on the -31
\kms\ map. 
}
\label{12m13cochannelspec}
\end{figure}

Maps of $^{13}$CO emission toward SNR G357.7+0.3 integrated over 4 velocity
ranges (+121$\pm$10, +102$\pm$10, +7$\pm$10 and -31$\pm$10
\kms) are shown in Figure \ref{12m13cochannelmap}. Each panel is labeled
in the top right with its velocity in red, with the location of peak as a
red square. The blue contours repeated on each panel are the non-thermal
radio brightness from the MOST Galactic Centre Survey \citep{gray94}.
Representative spectra from the locations of the bright peak of the 4
velocity maps (they are marked as squares in Figure
\ref{12m13cochannelmap}) are shown in Figure \ref{12m13cochannelspec}.  The
4 velocity components are likely at significantly different distance along
the line of sight, but their kinematic distances cannot be inferred from
the velocity and the galactic rotation curve, because the line of sight is
so close to the Galactic center. The +7 \kms\ component may be associated
with the eastern boundary of the SNR but has not been shown to be
associated. The supernova appears to be bounded by molecular gas in the -31
km~s$^{-1}$ component, in particular its southern, western, and
northwestern portions. The velocity of this component corresponds to that
of the shocked gas and in particular the narrow self-absorption toward the
broad CO. The red circle indicates the location of the OH 1720 MHz masers
which are at -35 to -37 km~s$^{-1}$. We already show broad molecular and
carbon line detection in its northwestern portions, and possible extended
interaction sites may be found to western and southern portions.

\begin{table*}
\caption[]{Observed Line brightness in the {\it Spitzer} IRS Spectra \label{tableirsline}}
\begin{center}
\begin{tabular}{llllllll}
\hline \hline
Wavelength &  Line & FWHM &  Line Brightness &  De-reddened Brightness\\
($\mu$m)     &       & ($\mu$m)  & (erg~s$^{-1}$~cm$^{-2}$~sr$^{-1}$) & (erg~s$^{-1}$~cm$^{-2}$~sr$^{-1}$) \\
\hline
    5.5004$\pm$    0.0016& H$_2$ S(7) &    0.056$\pm$    0.004& 2.29E-05$\pm$ 2.52E-06& 3.39E-05$\pm$ 3.74E-06\\
    6.9104$\pm$    0.0006& H$_2$ S(5) &    0.072$\pm$    0.002& 8.92E-05$\pm$ 2.44E-06& 1.21E-04$\pm$ 3.30E-06\\
    8.0037$\pm$    0.0007& H$_2$ S(4) &    0.061$\pm$    0.002& 3.52E-05$\pm$ 2.65E-06& 6.11E-05$\pm$ 4.59E-06\\
    9.6674$\pm$    0.0003& H$_2$ S(3) &    0.127$\pm$    0.001& 1.20E-04$\pm$ 9.82E-07& 5.71E-04$\pm$ 4.65E-06\\
   12.2821$\pm$    0.0002& H$_2$ S(2) &    0.082$\pm$    0.001& 6.13E-05$\pm$ 1.60E-06& 1.14E-04$\pm$ 2.97E-06\\
   17.0310$\pm$    0.0002& H$_2$ S(1) &    0.164$\pm$    0.001& 2.04E-04$\pm$ 8.03E-07& 3.46E-04$\pm$ 1.36E-06\\
   28.1932$\pm$    0.0012& H$_2$ S(0) &    0.299$\pm$    0.003& 3.08E-05$\pm$ 3.14E-07& 4.19E-05$\pm$ 4.28E-07\\
   34.8466$\pm$    0.0037& [Si~II]    &    0.196$\pm$    0.009& 3.19E-05$\pm$ 2.66E-06& 4.04E-05$\pm$ 3.37E-06\\
\hline \hline
\end{tabular}
\end{center}
\renewcommand{\baselinestretch}{0.8}
\end{table*}

\subsection{Molecular Hydrogen with {\it Spitzer}}

\begin{figure}
\includegraphics[width=9cm]{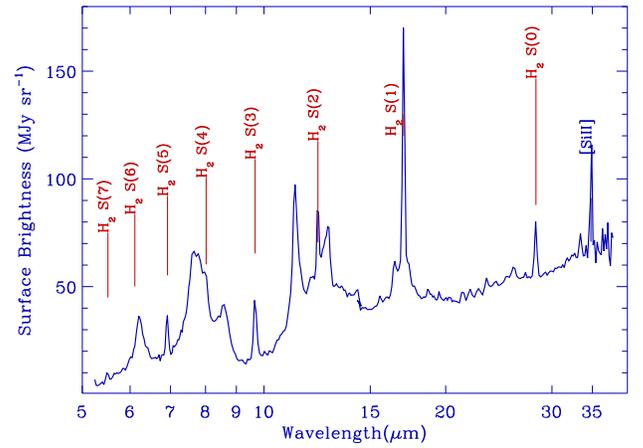}
\caption{{\it Spitzer} IRS spectrum of G357.7+0.3 showing strong H$_2$ but 
lacking ionic lines (only [Si~II] is shown).}
\label{g357h2}
\end{figure}

The {\it Spitzer} IRS spectrum where the broad CO emission from the SNR
G357.7+0.3 is detected as shown in Figure \ref{g357h2}.  All rotational 
H$_2$  lines within the IRS wavelength range are detected except S(6) line
at 6.109$\mu$m (note the feature appeared around 6.1$\mu$m is a part of PAH
emission and H$_2$ S(6) line is not detected; for comparison with those in
other SNRs, see Hewitt et al.\ 2011). The S(7) line is a weak detection.
The detected H$_2$ lines are S(0), S(1), S(2), S(3), S(4), S(5), and S(7)
as listed in Table \ref{tableirsline}. Interestingly, G357.7+0.3 shows a
significant lack of bright ionic lines compared with other SNRs which emit H$_2$
emission \citep{andersen11, hewitt09, neufeld06}. The only ionic line detected
is weak [Si II] at 34.8$\mu$m.

\begin{figure*}
\epsscale{0.85}
\includegraphics[scale=0.5,angle=0]{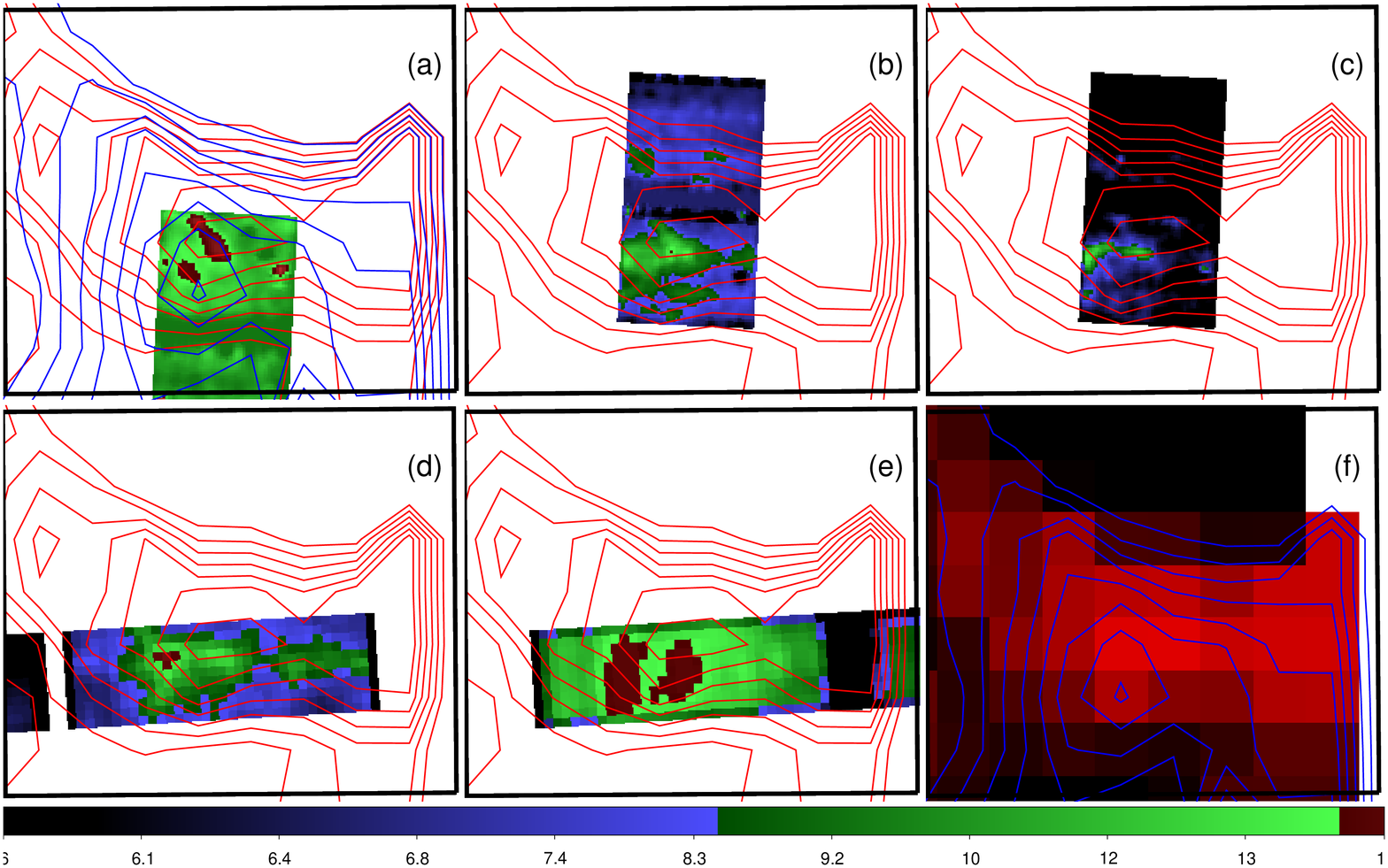}
\caption{H$_2$ maps of G357.7+0.3 at (a) S(5) at 6.9$\mu$m, (b) S(3) at 9.6$\mu$m,
(c) S(2) at 12.2$\mu$m, (d) S(1) at 17$\mu$m, (e) S(0) at 28$\mu$m, and (f)
CO(2-1) red wing of broad line image (in red) superposed on blue wing
contours. The region covered and contours is the same as in  Figure
\ref{cowing2}. The colorbar applies to panels (a)-(e) and the labeled
numbers apply to only panel (d) in units of MJy sr$^{-1}$. The FOV of the
images is 4.3$'$x3$'$ centered on R.A.\ $17^{\rm h} 38^{\rm m} 28^{\rm s}$
and Dec.\ $-30^\circ$ 34$^{\prime} 00^{\prime \prime}$.}
\label{g357h2maps}
\end{figure*}

The H$_2$ maps at different wavelengths are shown in Figure
\ref{g357h2maps}.  The H$_2$ map at 5.5$\mu$m is too weak to see any
structures, and the map at 8$\mu$m is blended with polycyclic aromatic
hydrocarbon (PAH) features. The H$_2$ maps show somewhat different
structures from each other. This is also seen in other regions such as HH
objects or other molecular interacting SNRs with shocked H$_2$ emission
\citep[][]{neufeld06, neufeld07}. The H$_2$ maps at 6.9 and 9.6$\mu$m (see
Figures \ref{g357h2maps}a and \ref{g357h2maps}b) both show a northwestern
rim along the CO blue wing emission and other H$_2$ maps at 12.2, 17 and
28$\mu$m (see Figures \ref{g357h2maps}c, \ref{g357h2maps}d and
\ref{g357h2maps}e) show elongated emission in the east-west direction,
around the CO peaks of red wing emission. The H$_2$ maps show a
concentration of emission at the peak of CO broad emission. However,
because of different spatial resolution between H$_2$ (3-6$''$) and CO maps
(beam size of 3$''$ vs. 22$''$), detailed structure could not be compared.
The one-to-one correspondence between H$_2$ emission and broad CO emission
is observed in IC 443 \citep[see][]{rho01}. H$_2$ emission from G357.7+0.3
is probably collisionally excited from a shock, but we can not rule out
contribution from UV pumping. Near-IR observations of vibrational H$_2$
lines are required to determine if there is contribution from UV pumping.
High resolution CO images such as using ALMA could produce CO maps with a
high spatial resolution comparable to or greater than those of H$_2$ images.

We have examined \spitzer\ IRAC and MIPS images (aorkey of 14296832 and 14658560),
and we did not find any emission associated with the SNR. IRAC band 2 emission
includes bright H$_2$ lines and shows shocked H$_2$ emission for many SNRs as shown
in the GLIMPSE survey \citep{reach06}. When we carefully examined IRAC band 2
emission, we note that the background emission (6-7 MJy sr$^{-1}$) is a factor of 2
-3 higher than those in other SNRs. The reason is likely because the background
emission near the Galactic Center is high. 

PAH emission appears in the spectrum of G357.7+0.3 shown in Figure
\ref{g357h2}. However, it is unclear if PAH emission belongs to the SNR or
not because the area covered by the IRS observations seem to be mostly
inside the SNR, except possibly for the SL1. We examined  H$_2$ map at
9.6$\mu$m where the northern part of IRS image may cover outside the SNR
(see Figure \ref{g357h2maps}b; note the boundary of the shock front is
unclear because of lack of high resolution images). When we assume the
emission in the northern part is background, the PAH emission disappeared.
When we examined mosaiked post-calibrated image (pbcd), we don't see much
variation of PAH emission. Many middle-aged SNRs show PAH emission as noted
by \citet{andersen11}. However, because the SNR is large compared to the
area we covered, future observations to cover a larger area including the
area outside the SNR are necessary to confirm or disprove that PAH emission
doesn't belong to the SNR G357.7+0.3.

We estimated the extinction value by using the silicate absorption dip around
10$\mu$m using PAHFIT (Smith et al. 2007). The fit yielded a value of optical depth
at 9.7$\mu$m, $\tau$(9.7$\mu$m), of 0.47, which is equivalent to an extinction value
of A$_v$ =  8.3 mag using an averaged value of Av/$\tau$(9.7)=18.5$\pm$2.0 from
observed values \citep{draine03}. This is equivalent to the line of sight column of
N$_H$ = 1.5$\times$10$^{22}$ cm$^{-2}$. This estimate is comparable to the value by
\citet{leahy89} assuming an average value of 3.$\times$10$^{21}$ cm$^{-2}$ per kpc.
The column density of G357.7+0.3 in the line of sight is comparable to that of W44
\citep{rho94}.



\begin{figure}
\plotone{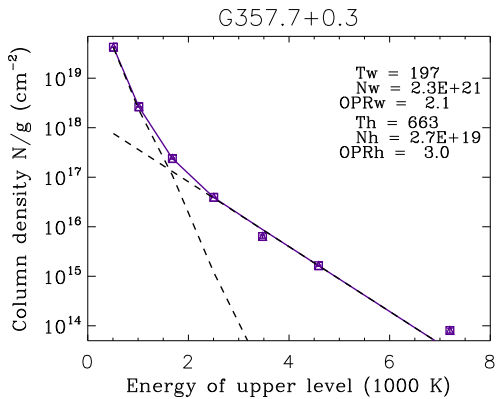}
\caption{H$_2$ excitation diagram of G357.7+0.3. The IRS data are marked as squares.
The fitting results of two-temperature LTE fit are shown.}
\label{g357h2ex}
\end{figure}

\begin{figure}
\includegraphics[scale=1.2,angle=90,width=9truecm,height=10truecm]{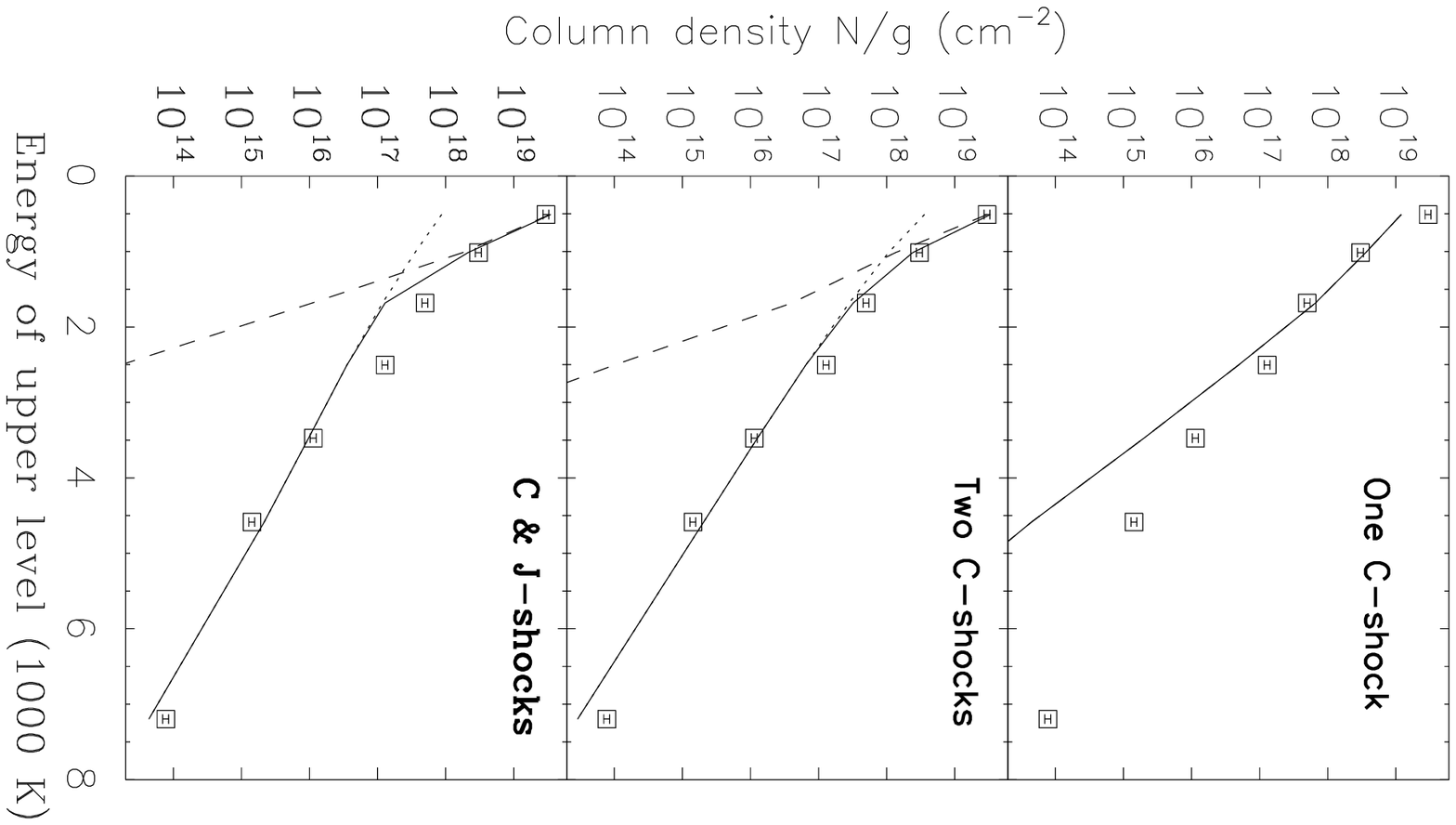}
\caption{Comparison of fitted shock models to excitation of H$_2$. The data are
shown with errors. From top to bottom, a single C-shock, two C-shocks and a
combination of one C-shock and one J-shock. The best fit results are listed
in Table \ref{Tshockmodel}. In cases of multiple-shocks, the slower shock
is plotted with a dashed line and the faster shock with a dotted line. The
total contribution of the two shock components is plotted as a solid line.}
\label{g357shockmodel}
\end{figure}


\section{Discussion}

\subsection{Excitation of Molecular Hydrogen}

The set of detected rotational H$_{2}$ lines in Table \ref{tableirsline} is an
excellent diagnostic of physical conditions in the shocked gas of
G357.7+0.3. An excitation diagram of the rotational H$_{2}$ lines was presented by
\citet{hewitt09}. An excitation diagram from the de-reddened brightnesses of H$_2$
lines is presented in Figure \ref{g357h2ex}. We have fit the data using a model of
two-temperature local-thermal equilibrium (LTE). The two-temperature fit yields a
warm temperature (T$_{\rm warm}$) of 197 K with a column density (N$_{\rm warm}$) of
2.3$\times$10$^{21}$ cm$^{-2}$ and an ortho-to-para ratio (OPR) of 2.1, and a high
temperature (T$_{hot}$) of 663 K with a column density (N$_{\rm hot}$) of
2.7$\times$10$^{19}$ cm$^{-2}$ and an OPR of 3.

The warm temperature (197 K) component shows an OPR of 2, indicating the
emission has not reached an equilibrium OPR. Conversion of para-to
ortho-H$_2$ behind the shock depends on collisions with atomic hydrogen and
is inefficient at low temperature due to the energy barrier (E/k
$\sim$4000) in that conversion. \citet{neufeld06} suggested that the
non-equilibrium H$_2$ OPRs were consistent with shock models in which the
gas is warm for a time period shorter than that required for reactive
collisions between H and para-H$_2$ to establish an equilibrium OPR. Hewitt
et al. (2009) show detection of H$_2$ from six SNRs and the H$_2$ emission
has a warm component with T$_w$ $\sim$250 - 550 K and a column density of
$\sim$10$^{20}$ cm$^{-2}$, and a hot component with T$_h$ $\sim$ 1000 -
2000 K and a column N$_w$$\sim$ 10$^{19}$ cm$^{-2}$. IC 443 shows a higher,
warm temperature of 627 K \citep{neufeld07} while G357.7+0.3 shows a lower
value 197 K for a warm temperature. There are differences between SNRs, and
non-LTE model may help to distinguish their difference in physical
conditions as the H$_2$ fitting is done with a simplified LTE model, H$_2$
S(7) line at 5.5$\mu$m in G357.7+0.3 is above the two-temperature model,
indicating there may be a third (hotter, $>$1000 K) component present as
other molecular SNRs show such hot temperature components
\citep{hewitt09,richter95,neufeld07}. However, the S(7) line in G357.7+0.3
is relatively weak compared with those in other molecular SNRs.
Near-infrared follow-up observations of H$_2$ lines combined with our {\it
Spitzer} rotational lines and non-LTE model may be required to identify
differences of physical condition of H$_2$ in G357.7+0.3. Nevertheless, we
find a difference in the best-fitted shock model in G357.7+0.3 from those
of other molecular SNRs as described below.

\subsection{Implication of Shock Models from Molecular Hydrogen}

We compare several shock models with the observed H$_2$ emission. We use
published models; C-shocks are from Le Bourlot et al. (2002), which was
applied to Orion and Wilgenbus et al. (2000), and J-shock models are from
\citet{hollenbach89}. We use a grid of shock  models and the H$_2$
excitation was fitted using least squares fitting; detailed methods were
described in \citet{hewitt09}. A grid of computed C-shock
models spans (log$_{10}$n$_{0}$) = 3, 4, 5, 6 cm$^{-3}$, v$_s$ = 10, 15,
20, 25, 30, 40 \kms, and OPR = 0.01, 1, 2, 3. A grid of J-shock models are
with densities of 10$^3$, 10$^4$, 10$^5$, 10$^6$ cm$^{-3}$ and shock
velocities of 30-150 \kms\ in 10 \kms\ increments. The fitted results are
shown in Figure \ref{g357shockmodel} and are summarized in Table
\ref{Tshockmodel}. The two C-shock model yielded the best model over one
component shock model or a combination of C-shock and J-shock models. The
best fit is a combination of two slow  C-shock models; a C-shock model with
a density of 10$^4$ cm$^{-3}$ and a velocity of 10\kms, and a second
component of C-shock with a density of 10$^5$ cm$^{-3}$ and the same
velocity of 10\kms. The errors of the estimated density,
shock velocity and OPR are limited to the available grid of the shock
models.

Other molecular SNRs generated an equivalent quality of the fitting  between a model
of two C-shocks and a combination of C- and J-shock models \citep[e.g.][]{hewitt09}.
In contrast, G357.7+0.3 strongly favors the two C-shock model over a combination of C-
and J-shocks based on H$_2$ model fitting. G357.7+0.3 also lacks ionic lines in the
IRS spectra; again in contrast to other molecular SNRs shown in \citet{andersen11} and
\citet{neufeld07}. Most importantly, the detection of [C~II] using SOFIA GREAT shows the
FWHM is only 16 \kms\, comparable to those of millimeter CO lines. All these facts
show evidence of C-shocks and against the presence of J-shocks. This is in 
contrast to the results in the SNR G349.7+0.2 which shows the presence of J-shocks by
showing a large line width ($\sim$150 \kms) of a molecular water line \citep{rho15}.

\begin{table} 
\caption[]{Summary of shock model fitting based on H$_2$ data \label{Tshockmodel}}
\begin{center}
\begin{tabular}{llllll}
\hline \hline
  Model    &  $\Delta\chi^2$ & density & velocity & OPR \\
           &  & (cm$^{-3}$) & (km s$^{-1}$) & &\\
\hline
 One C-shock  & 800  &  n=10$^3$   &  30           & 3 \\ 
\hline
{\bf Two C-shock}  & {\bf 17} & {\bf n=10$^4$}   & {\bf  30 (or 10)}   & {\bf 2}\\
              &  & {\bf  n=10$^5$}   &  {\bf 10}   & -- \\
\hline
 C-shock  & 148  & n=10$^3$   &  10   & --\\
   + J-shock  & & n=10$^5$   &  5   & 3\\
\hline
 C-shock  & 300 & n=10$^3$   &  20   & --\\
 + J-shock  &  & n=10$^6$   &  150   & --\\
\hline \hline
\end{tabular}
\end{center}
\end{table}

\subsection{Line Opacity of the CO Molecular Gas}

We use two pairs of $^{12}$CO and $^{13}$CO lines to estimate line opacities of
CO. The first pair is $^{12}$CO(3-2) and $^{13}$CO(3-2) lines and the second
pair is $^{12}$CO(2-1) and  $^{13}$CO(2-1) lines. The ratio of $^{12}$CO (3-2)
and $^{13}$CO(3-2) line intensities of the broad line is 17-40 where the ratio
varies depending on the velocity.  The broad line is defined as the lines at the
velocity between -58 and  -38 \kms\ (blue CO wing), and at the velocity between
-31 and  -27 \kms (red CO wing), and excludes the velocity range with the
self-absorption. The ratio between $^{12}$CO(2-1) and $^{13}$CO(2-1) is almost
the same as the ratio between $^{12}$CO(3-2) and $^{13}$CO(3-2) lines. 

The ratio of the $^{12}$CO/$^{13}$CO lines is related to the optical depth of the 
$^{12}$CO line, $\tau$, and the abundance ratio, $X$, of $^{12}$CO over $^{13}$CO: 
\begin{equation}
\frac{T_{12}}{T_{13}}  = \frac{1-e^{-\tau_{12}}}{1-e^{-\tau_{13}}} = \frac{1-e^{-\tau_{12}}}{1-e^{-{\tau_{12}/X}}} 
\end{equation}

We solved $\tau_{12}$ iteratively based on the
observed ratio of $T_{12}$/$T_{13}$ and we adopted
$X=60$ \citep{lucas98}. 
From the observed line ratio (17-40) of the broad lines, the optical depth
is in the range 0.9-3.6, which we observed is optically thin or slightly
optically thick gas.

We estimate the optical depth of the apparent absorption in the $^{12}$CO
lines (at the velocity between -38 and -31 \kms).
Estimating the center of the apparent absorption dip in intensity as a
factor of 3-17, the optical depth is 3.6-25. The cloud includes optically
thick gas which is the portion of the cold gas located in front of the
warm, shocked gas. The optical depth of CO is estimated for the molecular
clouds at velocities of 13 \kms\ in Figures \ref{smt12mspec} and
\ref{positionvelmap}. The ratio of the line brightnesses
$^{12}$CO/$^{13}$CO at 13 \kms\ is $\sim$5, which corresponds to the
optical depth of 13. The clouds at 13 \kms\ are optically thick from
unshocked gas and are not related to the SNR G357.7+0.3.

\begin{figure}
\includegraphics[scale=0.8,angle=90]{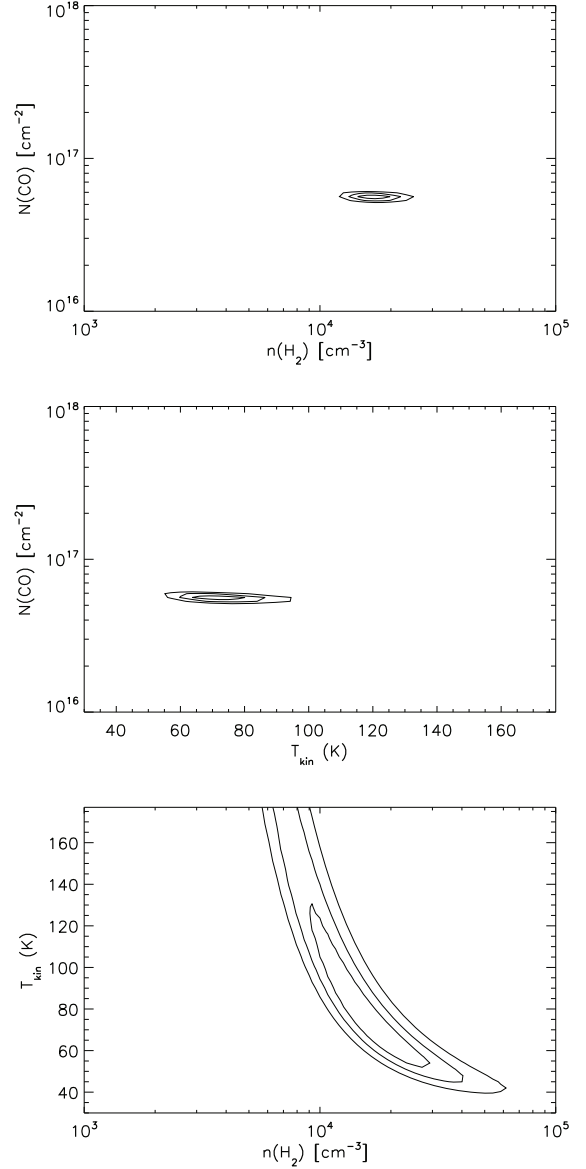}
\caption{Comparison of the CO line brightness to RADEX models at different H$_2$
density, $n({\rm H}_2)$, CO column density, $N({\rm CO})$, and gas temperature,
$T$. 
The contours are the goodness-of-fit, $\chi^2$, for models compared to all three line
brightnesses; contours are 99\%, 90\% and 67\% confidence intervals.}
\label{g357radexgrid}
\end{figure}

\begin{figure*}
\includegraphics[scale=0.84,angle=0]{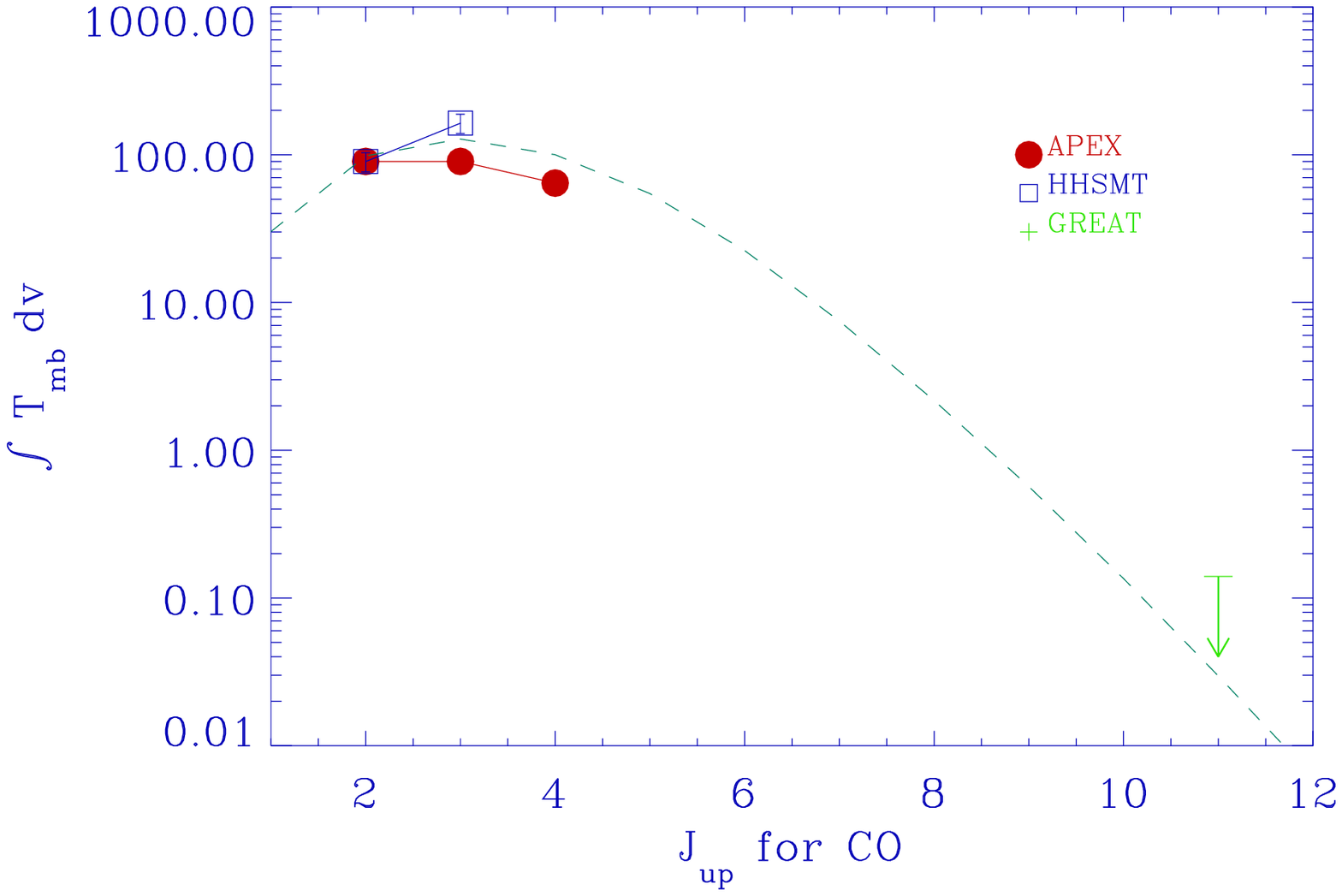}
\caption{CO surface brightness diagram as a function of J$_{up}$ with HHSMT and APEX
observations. The upper limit of SOFIA GREAT (green arrow) is shown. The
best-fit of RADEX model is shown in a dotted line (dark green).}
\label{g357cosbmodel}
\end{figure*}
\vskip 1truecm

\subsection{Large Velocity Gradient Analysis of the CO Molecular Gas}

Using the three observed line brightnesses of $^{12}$CO(2-1),
$^{12}$CO(3-2) and $^{12}$CO(4-3) in Table \ref{TCOlines}, we constrain the
physical conditions in the shocked CO gas. An average value between APEX
and HHSMT line intensities for the same line is used. We have made non-LTE
analysis using RADEX \citep{vandertak07}, which is a radiative transfer
code at the “intermediate” level; the most advanced methods that drop the
local approximation and solve for the intensities (or the radiative rates)
as functions of depth into the cloud, as well as of velocity. We use an
average velocity width of 17.3 \kms, and the line brightnesses calculated
from the averages of the measured line integrals divided by the line widths
from Table~\ref{TCOlines}. We assume that the emission is uniform on the
scale of the largest beam of 30$''$. Note that we don't have spatial
information on scales smaller than 30$''$ because we only have one spectrum
for each line except CO(2-1) and the beam size of CO(2-1) is 30$''$. Using
the RADEX model, confidence contours of a H$_2$ volume density density
[n(H$_2$)], CO column density [N(CO)], and gas kinetic temperature [T] are
obtained as shown in Figure \ref{g357radexgrid}. The best fit yields
n(H$_2$)=1.7$^{+0.8}_{-0.5}$$\times$10$^{4}$ cm$^{-3}$, N(CO) =
5.6$^{+0.1}_{-0.1}$$\times$10$^{16}$ cm$^{-2}$, and T=75$^{+30}_{-15}$ K.
The temperature inferred from the CO lines is roughly comparable to the
`warm' H$_2$ emission traced by the low-J H$_2$ rotational lines; if these
gases are from the same material, then the abundance of [CO/H$_2$] is
$2.4\times 10^{-5}$. That abundance is only 5\% of the maximum CO abundance
that would occur if all C and most O were locked into CO molecules
($5\times 10^{-4}$), which means much of the C is likely in atomic gas (as
C$^+$) or solids (contributing to the infrared continuum). Combining the
volume density inferred from the CO excitation with the H$_2$ column
density from the H$_2$ line brightness, the emitting region is 0.04$\pm
0.02$ pc along the line of sight. This short path length corresponds to
1$''$ on the sky, which is smaller than the resolutions of the telescopes,
despite the source appearing extended. When we examine the H$_2$ image at
6.9$\mu$m (which has the highest spatial resolution in all H$_2$ images) as
shown in Figure \ref{g357h2maps}, the size of the smallest knot is about
3$''$. This is the limit of spatial resolution of the IRAC image. Thus, the
1$''$ structures were not resolved by our H$_2$ images. The likely geometry
of the emitting region is thin sheets, which are the post-shock regions
with shock fronts spanning regions large than the beam.

Figure \ref{g357cosbmodel} shows the CO surface brightness as a function of
upper rotation level (J$_{upper}$) of CO. The best fit model is shown as a
dotted line. The upper limit of SOFIA CO(11-10) line is above the
brightness from the model and the model indicates that a longer observation
with SOFIA GREAT would have detected the line (note that the integration
time of CO(11-10) was just 5 mininues). The volume density, n(H$_2$), or
the emitting CO gas in G357.7+0.3 is lower than those of a few SNRs (which
includes a density of 10$^{6}$ cm$^{-3}$) \citep[for example][]{hewitt09},
but comparable to that of W28 \citep{neufeld14}.

The critical density of CO(4-3) is 3.2$\times$10$^{4}$
(3.7$\times$10$^{4}$, 2.4$\times$10$^{4}$) cm$^{-3}$ for 200 K (30, 3000
K), and CO(3-2) is 1.03$\times$10$^{4}$ (1.1$\times$10$^{4}$,
8.6$\times$10$^{3}$) cm$^{-3}$) for 200 K (30, 3000 K), and CO(2-1) is
2.04$\times$10$^{3}$ (2.2$\times$10$^{3}$, 1.9$\times$10$^{3}$) cm$^{-3}$
for 200 K (30, 3000 K), respectively. Because the critical density of CO is
low, CO is often used as a thermometer of the ISM. The volume density
derived by a RADEX model is slightly lower than the critical densities of
CO(4-3) and CO(3-2) and higher than that of CO(2-1). Shocked gas of CO(2-1)
is collisionally dominated while the CO gas emitting CO (4-3) and CO(3-2)
is partially subthermal.

\section{Conclusion}

1. From the relatively unknown SNR G357.7+0.3, we discover broad molecular lines of
CO(2-1), CO(3-2), CO(4-3), $^{13}$CO (2-1) and $^{13}$CO (3-2), HCO$^+$ and HCN
using the HHSMT, 12-Meter Telescope, APEX and MOPRA
telescopes. The widths of the broad lines are 15-30 \kms, that are caused by
strong supernova (SN) shocks passing through dense molecular clouds.
The detection of such broad lines is unambiguous, direct evidence of shocked gas.
This is the first evidence showing that G357.7+0.3 is a SNR interacting with
molecular clouds. 

2. We present detection of shocked molecular hydrogen (H$_2$) in the mid-infrared
using the {\it Spitzer} IRS observations. The observations covered an area of
about 1$'$ with short-low and $\sim$3$'$$\times$1$'$ with long-low. The
rotational H$_2$ lines of S(0)-S(5), and S(7) with the IRS are detected. The
detection of H$_2$ lines is also evidence that G357.7+0.3 is interacting with
molecular clouds. The two-temperature LTE fit yields a warm temperature (T$_{\rm
warm}$) of 197 K with a column density (N$_{\rm warm}$) of 2.3$\times$10$^{21}$
cm$^{-2}$ and an ortho-to-para ratio (OPR) of 2.1, and a high temperature
(T$_{hot}$) of 663 K with a column density (N$_{\rm hot}$) of
2.7$\times$10$^{19}$ cm$^{-2}$ and an OPR of 3. The ortho-to-para ratio of the
low temperature component is less than 3, indicating that the SNR G357.7+0.3 is 
propagating into cold quiescent clouds.

4. We observed [C~II] at 158$\mu$m and high-J CO(11-10) observations with the
GREAT on board SOFIA. CO(11-10) is not detected, but GREAT spectrum of
[C~II] shows a 3$\sigma$ detection, with  a broad line of a width of 15.7
\kms\ that had  a line profile similar to those of millimeter CO lines. The
line width of [C~II] implies that ionic lines can come from a low-velocity
C-shock.

5. We have mapped a large-scale molecular cloud structure surrounding the
SNR G357.7+0.3 using the $^{13}$CO(1-0) line that typically traces cold
clouds. The supernova component at -31 km~s$^{-1}$ (integrated over -21 to
-41 \kms) appears to be bounded by molecular gas that is located at its
southern, western, and northwestern portions. We show broad molecular and
carbon line detections in its northwestern portions, and possible extended
interaction sites may be found toward western and southern portions with
future observations. 

6.  We compare shock models with the observed H$_2$ emission. A two C-shock model
yielded the best fit over a one component shock model or a combination of
C-shock and J-shock models. The best fit model of two slow  C-shock models; a
C-shock model with a density of 10$^4$ cm$^{-3}$ and a velocity of 10\kms, and a
second component of C-shock with a density of 10$^5$ cm$^{-3}$ and the same
velocity of 10\kms. G357.7+0.3 also lacks ionic lines in the IRS spectra. Most
importantly, the detection of [C~II] using SOFIA GREAT shows the FWHM is
$\sim$16 \kms\, comparable to those of millimeter CO lines. All these facts show
evidence of C-shocks and against the presence of J-shocks.

7. We estimate the CO density, column density, and temperature by running
RADEX models, using an average velocity
width of 17.3 km/s. The best fit yields
n(H$_2$)=1.7$^{+0.8}_{-0.5}$$\times$10$^{4}$ cm$^{-3}$, N(CO) =
5.6$^{+0.1}_{-0.1}$$\times$10$^{16}$ cm$^{-2}$, T=75$^{+30}_{-15}$ K. This
model is consistent with the upper limit of CO(11-10) brightness.

G357.7+0.3 shows broad CO lines for 4.5$'$$\times$5$'$ area and the broad lines
may extend to NE and to SW, beyond the area covered by our observations. The
interaction area showing CO broad lines and H$_2$ emission is large so the
pattern of the molecular cloud interaction with the SNR may be similar to those
of the well-known molecular SNRs of IC 443 and W44. It would be worthwhile to
extend the millimeter maps such as CO and infrared maps in H$_2$ which would
reveal the entire regions of interaction between the SNR and molecular clouds.
The newly discovered molecular cloud interaction with SNR G357.7+0.3 offers many exciting
opportunities of astrophysical laboratory to study dynamics of shocks, molecular
astro-chemistry, and high-energy phenomena in shocks and dense environment.

\acknowledgements

We thank Sebastien Bardeau, a staff scientist at IRAM for helping with various
issues of CLASS softwares, and Miguel Angel Requena Torres and Friedrich Wyrowski 
for helping with SOFIA and APEX observations and data processing, respectively. 
We thank anonymous referee for helpful comments. 
The Arizona Radio Observatory is part of the Steward
Observatory at the University of Arizona and receives partial support from the
National Science Foundation. Based [in part] on observations made with the NASA/DLR
Stratospheric Observatory for Infrared Astronomy. SOFIA Science Mission Operations
are conducted jointly by the Universities Space Research Association, Inc., under
NASA contract NAS2-97001, and the Deutsches SOFIA Institut under DLR contract 50 OK
0901. 
APEX is a collaboration between the Max-Planck-Institut f\"{u}r Radioastronomie, the European 
Southern Observatory, and the Onsala Space Observatory.

{}


\begin{thebibliography}{}
 
\bibitem[Abdo et al.(2009)]{abdo09} Abdo, A. A., Ackermann, M., Ajello, M., Baldini, L., Ballet, J.,
     Barbiellini, G., Baring, M. G., Bastieri, D. et al. 2009, \apj, 706, L1
\bibitem[Abdo et al.(2010a)]{abdo10a} Abdo, A. A. et al. 2010a, \apj, 712, 459 (IC 443)
\bibitem[Abdo et al.(2010b)]{abdo10c} Abdo, A. A. et al. 2010b, Science, 327, 1103 (W44)
\bibitem[Abdo et al.(2010c)]{abdo10b} Abdo, A. A. et al. 2010c, \apj, 718, 348 (W28)
\bibitem[Aharonian et al.(2008)]{aharonian08} Aharonian, F. et al., 2008, \aap, 490, 685
\bibitem[Anderl, Gusdorf, \& G\"{u}sten (2014)]{anderl14} Anderl, S., Gusdorf, A., \& G\"{u}sten, R., 2014, A\&A, 569, 81
\bibitem[Andersen et al(2011)]{andersen11}Andersen, M., Rho, J., Reach, W. T., Hewitt, J. W., \& Bernard, J. P., 2011, ApJ, 742, 7

\bibitem[Arikawa et al.(1999)]{arikawa99}Arikawa, Y., Tatematsu, K., Sekimoto, Y., \& Takahashi, T. 1999, PASJ, 51, L7
\bibitem[Barber et al.(2006)]{barber06}Barber, R. J., Tennyson, J., Harris, G. J., \& Tolchenov, R. N., 2006, MNRAS, 368, 1087
\bibitem[Burton et al.(1990)]{burton90} Burton, M. G., Hollenbach, D. J., Haas, M. R., Erickson, E. F., 1990, ApJ, 355, 197
\bibitem[Burton et al.(1992)]{burton92} Burton, M. G. \&  Hollenbach, D. J., 1992, ApJ, 399, 563
\bibitem[Chevalier (1999)]{chevalier99} Chevalier, R. A., 1999, 511, 798
\bibitem[Daniel \& Slane (2010)]{daniel10}Daniel, C. \& Slane, P., 2010, ApJ, 717, 372
\bibitem[Draine, Roberge \& Dalgarno(1983)]{draine83} Draine, B. T., Roberge, W. G. \& Dalgarno, A., 1983, ApJ, 264, 485
\bibitem[Draine(2003)]{draine03} Draine, B. T., 2003, ARA\&A, 41, 241
\bibitem[Dubner et al. (2004)]{dubner04}Dubner, G., Giacani, E., Reynoso, E., Par\'{o}n, S.,2004, A\&A, 426, 201
\bibitem[Esposito et al.(1996)]{esposito96}Esposito, J.A., Hunters, S.D., Kanbach, G., \& Sreekumar, P., 1996, ApJ, 461, 820
\bibitem[Faure \& Josselin (2008)]{faure08} Faure, A. \& Josselin, E., 2008, A\&A, 492, 257
\bibitem[Flower \& Pineau Des For\'{e}ts (2010)]{flower10} Flower, D. R. \& Pineau Des For\'{e}ts G., 2010, MNRAS, 406, 1745
\bibitem[Frail et al.(1996)]{frail96}Frail, D. A. et al. 1996, AJ, 111, 1651
\bibitem[Gray (1994)]{gray94}Gray, A.D, 1994, MNRAS, 270, 835
\bibitem[Hanabata et al.(2014)]{hanabata14} Hanabata, Y., Katagiri, H., Hewitt, J. W. et al., 2014, ApJ, 786, 145
\bibitem[Heyminck et al.(2012)]{heyminck12} Heyminck, S., Grad, U.U., G\"{u}sten, et al, 2012, A\&A, 542, L1 
\bibitem[Hewitt et al.(2009)]{hewitt09} Hewitt, J. W., Rho, J., Andersen, M., \& Reach, W. T., 2009, ApJ, 694, 1266
\bibitem[Hewitt et al.(2008)]{hewitt08} Hewitt, J. W., Yusef-Zadeh, F., Wardle, M., 2008, ApJ, 683, 189
\bibitem[Hewitt et al.(2012)]{hewitt12}Hewitt, J. W. et al., 2012, 759, 89
\bibitem[Hewitt et al.(2015)]{hewitt15}Hewitt, J. W. \& Lemoine-Goumard, M. 2015, {\it High-energy gamma-ray astronomy}, eds. B. Degrange, G. Fontaine, vol.1 
\bibitem[Hollenbach \& McKee(1989)]{hollenbach89} Hollenbach, D. \& McKee, C. F., ApJ, 1989, 342, 306 (HM89)
\bibitem[Hollenbach et al.(2009)]{hollenbach09}Hollenbach, D. Kaufman, M.J, Bergin, E. A., 
\& Melnick, G.J., 2009, ApJ, 690, 1497
\bibitem[Huang \& Thaddeus(1986)]{huang86} Huang, Y.-L. \& Thaddeus, P., 1986, ApJ, 309, 804
\bibitem[Jeong et al.(2013)]{jeong13}Jeong, I.-G. et al., 2013, ApJ, 770, 105
\bibitem[Kaufman \& Neufeld (1996)]{kaufman96}Kaufman, M.J, \& Neufeld, D. A. 1996, ApJ, 456, 611
\bibitem[Kawasaki et al.(2005)]{kawasaki05} Kawasaki, J. et al., 2005, ApJ, 631, 935

\bibitem[Kilpatrick et al.(2014)]{kilpatrick14} Kilpatrick, C.D., Bieging, J.H., \& Rieke, G.H. 2014, ApJ, 796, 144
\bibitem[Kilpatrick et al.(2016)]{kilpatrick16} Kilpatrick, C.D., Bieging, J.H., \& Rieke, G.H. 2016, ApJ, 816, 1 (scheduled) or arXiv:1511.03318
\bibitem[Kristensen et al.(2010)]{kristensen10} Kristensen et al. 2010, A\& 521, L30
\bibitem[Kristensen et al.(2012)]{kristensen12} Kristensen et al. 2012, A\& 542, 8
\bibitem[Koo \& Moon(1997)]{koo97} Koo, B.-C., \& Moon, D.-S., 1997, ApJ, 485, 263
\bibitem[Koo et al(2001)]{koo01}Koo, B.-C., Rho, J., Reach, W. T., Jung, J., \& Mangum, J. G., 2001, ApJ, 552,175
\bibitem[Ladd et al.(2005)]{ladd05} Ladd, N., Purchell, C., Wong, T., \& Robertson, S., 2005, PASA, 2005, 22, 62
\bibitem[Lazendic et al.(2005)]{lazendic05} Lazendic, J. S., Slane, P. O.,  Hughes, J. P., Chen, Y., 
Dame, T. M., 2005, 618, 734
\bibitem[Lazendic et al.(2010)]{lazendic10} Lazendic, J. S., Wardle, M., Whiteoak, J.B., Burton, M. G., \& Green, A.J., 2010, MNRAS, 409, 371
\bibitem[Leahy(1989)]{leahy89} Leahy, D. A., 1989, A\&A, 216, 193
\bibitem[Lesaffre et al.(2004a)]{lesaffre04a}
Lesaffre, P., Chi$\acute{e}$ze, J.-P., Cabrit, S., \& Pineau des For$\acute{e}$ts, G., 2004a, A\&A, 427, 147
\bibitem[Lesaffre et al.(2004b)]{lesaffre04b}
Lesaffre, P., Chi$\acute{e}$ze, J.-P., Cabrit, S., \&Pineau des For$\acute{e}$ts, G., 2004b, A\&A, 427, 157
\bibitem[Lucas \& Liszt(1998)]{lucas98}Liszt, H. S. \& Lucas, R., 1998, 337, 246 
\bibitem[Liszt \& Lucas(1998)]{liszt98}Liszt, H. S. \& Lucas, R., 1998, 339, 561
\bibitem[Liszt (2009)]{liszt09}Liszt, H. S., 2009, A\&A, 508, 1331L
\bibitem[Neufeld et al.(2006)]{neufeld06} Neufeld, D. A. et al., 2006, ApJ, 649, 816
\bibitem[Neufeld et al.(2007)]{neufeld07} Neufeld, D. A. et al., 2007, ApJ, 664, 890
\bibitem[Neufeld et al.(2014)]{neufeld14} Neufeld, D. A., Gusdorf, A., G\"{u}sten, Rolf; 
Herczeg, G. J., Kristensen, L., Melnick, Gary, J., Nisini, B., Ossenkopf, V., Tafalla, M., van Dishoeck, E. 781, 102
\bibitem[Orlando et al.(2009)]{orlando09} Orlando, S. Bocchino, F., Miceli, M., Reale, F., \& Peres, Mem. S.A.It. suppl. 13, 97\\
\bibitem[Pannuti et al.(2014)]{pannuti14} Pannuti, T.G., Rho, J., Heinke, C. O. \& Moffitt, W. P., 2014, AJ, 147, 55
\bibitem[Phillips et al.(2009)]{phillips09} Phillips, J. P., Ramos-Larios, G., \& Perez-Grana, J.A., 2009, MNRAS, 397, 1215
\bibitem[Phillips \& Marquez-Lugo(2010)]{phillips10} Phillips, J. P. \& Marquez-Lugo, R.A., 2010, MNRAS, 409, 701
\bibitem[Reach \& Rho (1998)]{reach98} Reach, W.T., and Rho, J., 1998, ApJ, 507, 93L
\bibitem[Reach \& Rho (1999)]{reach99} Reach, W.T., and Rho, J., 1999, ApJ,511, 836 
\bibitem[Reach \& Rho (2000)]{reach00}Reach, W. T., \& Rho, J.-H. 2000, ApJ, 544, 843
\bibitem[Reach et al.(2006)]{reach06}Reach, W. T., Rho, J., Tappe, A. et al., 2006, AJ, 131, 1479,
\bibitem[Reich \& F\"{u}rst(1984)]{reich84} Reich, W. \& F\"{u}rst, E., 1984, A\&AS, 57, 165
\bibitem[Rho et al.(2015)]{rho15} Rho, J., Hewitt, J. W., Boogert, A., Kaufman, M. \& Gusdorf, A., 2015, ApJ, 812, 44
\bibitem[Rho et al.(2001)]{rho01} Rho, J., Jarrett, T., Reach, W. T., \& Cutri,  
R., 2001, ApJ,547,885 
\bibitem[Rho \& Petre(1998)]{rho98} Rho, J. \& Petre, R., 1998, ApJ, 503, L167 
\bibitem[Rho et al.(1994)]{rho94} Rho, J., Petre, R., Schlegel, E. M., and Hester, J.J., 1994, ApJ, 430, 757
\bibitem[Reach \& Rho (1999))]{reach99}Reach, W. T., Rho, J., 1999, 511, 836
\bibitem[Reach, Rho \& Jarrett(2005)]{reach05}Reach, W. T., Rho, J., Jarrett, T. H., 2005, ApJ, 618, 297
\bibitem[Richter, Graham \& Write(1995)]{richter95}Richter, M.J, Graham, J.R., \& Wright, G. S., 1995, ApJ, 454, 277
\bibitem[Roelfsema et al.(2012)]{roelfsema12}Roelfsema, P.R. et al., 2012, A\&A, 537, 17
\bibitem[Seta et al.(2004)]{seta04}Seta, M., Hasegawa, T., Sakamoto, S., Oka, T., Sawada, T., Insutsuka, S.,
Koyama, H., \& Hayashi, M. 2004, AJ, 127, 1098
\bibitem[Sezer et al.(2011)]{sezer11} Sezer, A., G\"{o}k, F., Hudaverdi, M. \& Ercan, E. N.
    2011a, \mnras, 417, 1387
\bibitem[Shelton et al.(2004)]{shelton04}Shelton, R. L., Kuntz, K. D., \& Petre, R. 2004, \apj, 615, 275
\bibitem[Shinn et al.(2010)]{shinn10}Shinn, J.-H. et al.,2010, AdSpR, 45, 445
\bibitem[Slane et al(2002)]{slane02}  Slane, P., Chen, Y., Lazendic, J. S., Hughes, J. 2002, ApJ, 580, 904
\bibitem[Smith \& Draine(2007)]{smith07} Smith, J.D.T., Draine B.T., et al., 2007, ApJ, 656, 770
\bibitem[Snell et al(2005)]{snell05} Snell, R.L., Hollenbach, D., Howe, J.E., Neufeld, D. A.,
Kaufman, J.J., Melnick, G.J., 2005, ApJ, 620, 758
\bibitem[Sezer et al.(2011)]{Sezer11a} Sezer, A., G\"{o}k, F., Hudaverdi, M. \& Ercan, E. N.
    2011a, \mnras, 417, 1387
\bibitem[Tennyson et al.(2001)]{tennyson01} Tennyson, J., Zobov, N.F., Williamson, R., Polyansky, O. L., \& Bemath, P. F., P2001, J. Phys. Chem. Ref. Data, 30, 735
\bibitem[Tilley et al.(2006a)]{tilley06a}Tilley, D.A., \&  Balsara, D.S., 2006a, ApJL, 645, 49
\bibitem[Tilley et al.(2006b)]{tilley06b}Tilley, D.A., Balsara, D.S. \& Howk, J.C. 2006b, \mnras, 371, 1106
\bibitem[Uchiyama et al.(2010)]{uchiyama10} Uchiyama, Y., Blandford, R. D., Funk, S., Tajima, H. \&
    Tanaka, T. 2010, \apj, 723, L122
\bibitem[van Dishoeck et al.(1993)]{vandishoeck93}van Dishoeck, E. F., Jansen, D.J, \& Phillips, T.G., 1993, A\&A, 279,541
\bibitem[van Dishoeck et al.(2011)]{vandishoeck11}van Dishoeck, E. F. et al., PASP, 123, 138
\bibitem[van der Tak et al.(2007)]{vandertak07}van der Tak, F.F.S., Black, J. H. Sch\"{o}ier, F.L, Jansen, D.J., van Dishoeck, E. F., A\&A, 468, 627
\bibitem[Wootten (1997)]{wootten77}Wootten, H. A., 1977, ApJ, 216, 440
\bibitem[Wu et al.(2011)]{wu11}Wu, J. H. K. et al., 2011, ApJL, 740, L12
\bibitem[Yamauchi et al.(1998)]{yamauchi98}Yamauchi, S., Koyama, K., Kinugasa, K., Torii, K., Nishiuchi, M., Kosuga, T.,
 Kamata, Y., 1998, Astron. Nachr, 319,111
\bibitem[Young et al.(2012)]{young12} Young, E.T., Becklin, E.E., Marcum, P. M. et al. 2012, ApJ, 749, L17
\bibitem[Yusef-Zadeh et al.(1999)]{yusef-zadeh99} Yusef-Zadeh, F., Goss, W.M., Roberts, D. A., Robinson, B., \& Frail, D. A.,  ApJ, 527, 172
\bibitem[Zhou et al.(2011)]{zhou11}Zhou, X. et al., 2011, ApJ, 791, 109
\bibitem[Zhou et al.(2014)]{zhou14}Zhou, X. et al., 2014, ApJ, 743, 4
\end{thebibliography}
\end{document}